\documentclass[aps,pra,reprint,groupedaddress]{revtex4-2}
\usepackage{graphicx}
\usepackage{epstopdf}
\graphicspath{figures/}
\usepackage{dcolumn}
\usepackage{amssymb}
\usepackage{amsmath}
\usepackage{bm}
\usepackage{times}
\usepackage{mhchem}
\usepackage{lineno}
\usepackage{makecell}
\usepackage[colorlinks,linkcolor=blue,anchorcolor=blue,citecolor=blue,urlcolor=blue]{hyperref}
\UseRawInputEncoding
\begin{document}
    \title{Nonlinear Tripartite Coupling of Trapped Electrons with Magnons in a Hybrid Quantum System}
    \author{Xue-Feng Pan}
    \author{Peng-Bo Li}
    \email{lipengbo@mail.xjtu.edu.cn}
    \affiliation{Ministry of Education Key Laboratory for Nonequilibrium Synthesis and Modulation of Condensed Matter, Shaanxi Province Key Laboratory of Quantum Information and Quantum Optoelectronic Devices, School of Physics, Xi'an Jiaotong University, Xi'an 710049, China}

   \date{\today}

\begin{abstract}
Coherent nonlinear tripartite interactions are critical for advancing quantum simulation and information processing in hybrid quantum systems, yet they remain experimentally challenging and still evade comprehensive exploration. Here, we predict a nonlinear tripartite coupling mechanism in a hybrid setup comprising a single trapped electron and a nearby micromagnet. The tripartite coupling here leverages the electron's intrinsic charge (motional) and spin degrees of freedom interacting with the magnon modes of the micromagnet. Thanks to the large spatial extent of the electron zero-point motion, we show that it is possible to obtain a tunable and strong spin-magnon-motion coupling at the single quantum level, with two phonons simultaneously interacting with a single spin and magnon excitation.
This enables, for example, magnons to mediate coupling among distinct degrees of freedom of two electrons, which can be used for the rapid preparation of few-body entangled states.
This protocol can be readily implemented with the well-developed techniques in electron traps and quantum magnonics, and may open new avenues for quantum simulations and hybrid quantum information processing by introducing a versatile platform for exploring multipartite interactions and nonclassical state generation.
\end{abstract}
\maketitle

\section{Introduction}
Coherent interactions
between degrees of freedom of completely different nature
are the foundation
for quantum information processing with hybrid
quantum systems \cite{2013XiangP623653,2020BurkardP129140,2020ClerkP257267} and have been widely used to
explore new quantum applications like quantum simulations~\cite{2025MaskaraP289297,2022DaleyP667676,2014GeorgescuP153185,2021MonroeP2500125001,2024SoP80118011,2017GrossP9951001}.
Remarkably,  strongly coupled hybrid quantum systems based on magnons in micromagnets with other quantum systems~\cite{2016TabuchiP729739,2019LachanceQuirionP7010170101,2021LiP164,2021YuP159,2025YuanP2442224422,2024ZuoP3120131201,2022WangP75807580,2024PanP1302013020,2024FlebusP363501363501,2024YuP186,2018PirmoradianP224409224409}, including solid-state spins~\cite{2020NeumanP247702247702,2023HeiP7360273602,2021HeiP4370643706,2022XiongP245310245310,2023JiP180409180409,2024FukamiP23137541202313754120,2021FukamiP4031440314,2017AndrichP2828,2020CandidoP1100111001,2021SkogvollP6400864008,2022GonzalezBallesteroP7541075410,2024BejaranoP20422042,2013TrifunovicP4102341023}, photons~\cite{2018HaighP214423214423,2014ZhangP156401156401,2022KaniP257201257201,2020YuP107202107202,2020YangP147202147202,2024BinP4360143601,2018HarderP137203137203,2024KimP90149014,2021LuP6370863708,2024LiuP2370923709,2018OsadaP103018103018,2023QianP34373437,2023YuanP134409134409,2023HanP47304730,2023PanP1407514075,2023YangP5441354413,2025LiP1351913519,2024QianP156901156901,2024UllahP6371563715,2022PanP5442554425,2022RameshtiP161,2022BittencourtP183603183603,2020YuanP5360253602,2025YuanP1443314433,2010SoykalP7720277202,2014TabuchiP8360383603,2014TabuchiP8360383603,2015ZhangP1501415014,2016WangP224410224410,2017SharmaP9441294412,2018WangP5720257202,2018WangP5720257202,2019WangP127202127202,2015LambertP5391053910,2019KongP3400134001,2016KostylevP6240262402,2018OsadaP133602133602,2019LiP2180121801,2023AsjadP37,2022KaniP1360213602,2022ShenP243601243601,2025WangP1370913709,2022ShenP123601123601}, phonons~\cite{2022ShenP123601123601,2025WangP1370913709,2022ShenP243601243601,2022KaniP1360213602,2023AsjadP37,2019LiP2180121801,2016ZhangP15012861501286,2018LiP203601203601,2020GonzalezBallesteroP125404125404,2020GonzalezBallesteroP9360293602,2020ColombanoP147201147201,2021LiP6090260902,2023ZengP1300313003,2023WangP110311110311,2023ChengP4319743197,2025HeP24002752400275,2024CarraraP216701216701}, skyrmion qubits~\cite{2024PanP193601193601}, and superconducting qubits~\cite{2015TabuchiP405408,2020WolskiP117701117701,2022KounalakisP3720537205,2023XuP193603193603,2023JinP5370253702,2020XieP4233142331,2019LiuP134421134421,2024DolsP104416104416,2023KounalakisP224416224416,2024HouP1371113711}, have been extensively investigated.
In such configurations, the micromagnet acts as a microwave nanomagnonic cavity capable of confining magnonic excitations within deep subwavelength spatial domains.
All these make quantum magnons particularly attractive as promising carriers of quantum information in emerging hybrid quantum technologies~\cite{2014KrawczykP123202123202,2015ChumakP453461,2021PirroP11141135,2022YuanP174,2021LiP4034440344,2023ChenP2410524105,2022YangP1241912419,2024YangP206902206902,2024ZouP3670136701,2019RusconiP2234322343,2024HeiP4402544025}.

Recently, electrons trapped above the surface of liquid He (eHe qubit) and solid Ne (eNe qubit) have attracted great interest in the field of quantum
science and technology~\cite{2024JenningsP120501120501,2024GuoP240615870,2020KjaergaardP369395,1999PlatzmanP19671969,2000NietoP3490134901,2003DykmanP155402155402,2010MonarkhaP565575,
2019KawakamiP8680186801,2021KawakamiP106802106802,
2024XieP5260752607,2022ZhouP4650,2024ZhouP116122,2016YangP1103111031,2019KoolstraP53235323,2010SchusterP4050340503,2024KanaiP250603250603,2009ZhangP5580155801,2024BeysengulovP3032430324,2023KawakamiP5402254022,2012ZhangP205408205408,2023DykmanP3543735437,
2006LyonP5233852338,2022ChenP4501645016,2011BradburyP266803266803,2005ZavyalovP415420,2020JinP3500335003,2025LiP250720476,2025LiP10052502}.
Compared to traditional charge and superconducting qubits, these kinds of artificial atoms overcome the limited coherence times caused by material defects and background noise in charge qubits, as well as the short coherence times and size reduction challenges in superconducting qubits~\cite{2024JenningsP120501120501,2020KjaergaardP369395}.
In addition, the trapped electron can form both a charge qubit (based on its quantized motional states)~\cite{2009ZhangP5580155801,2010SchusterP4050340503,2016YangP1103111031,2019KoolstraP53235323,2024BeysengulovP3032430324,2022ZhouP4650,2024KanaiP250603250603,2024ZhouP116122} and a spin qubit (utilizing its spin degree of freedom)~\cite{2006LyonP5233852338,2012ZhangP205408205408,2023DykmanP3543735437,2022ChenP4501645016}, each of which can be precisely controlled and dynamically tuned through electric or magnetic fields.

Hybrid quantum systems employing eNe and eHe qubits are predominantly realized within the circuit quantum electrodynamics (cQED) framework, incorporating coupling to electronic motional states~\cite{2022ZhouP4650,2024ZhouP116122,2010SchusterP4050340503,2016YangP1103111031,2019KoolstraP53235323}.
In such systems, the in-plane quantized motion can be engineered with transition frequencies on the order of several GHz, enabling coherent interaction with on-chip resonant cavities.
Previous studies have proposed motion-spin coupling via magnetic field gradients~\cite{2010SchusterP4050340503,2012ZhangP205408205408,2023DykmanP3543735437,2023KawakamiP5402254022} and have also explored photon-spin interactions mediated by motional states~\cite{2010SchusterP4050340503} as well as electron-electron coupling through Coulomb interactions~\cite{2023DykmanP3543735437,2024BeysengulovP3032430324,2025LiP250323738}.
However, the simultaneous realization of \textit{direct tripartite coupling} remains a significant challenge in these schemes.
In magnonic systems, magnons generate spatially varying quantized magnetic fields, enabling direct tripartite coupling with phonons and spins~\cite{2023HeiP7360273602}.
However, in previous studies, the extremely small zero-point fluctuations of phonon associated motional states (on the femtometer scale) have rendered higher-order contributions in tripartite interactions negligible.
Consequently, how to effectively enhance \textit{higher-order tripartite couplings} to render their experimental investigation feasible has become an urgent scientific problem.

To address this challenge, we propose a hybrid quantum device comprising a yttrium iron garnet (YIG) micromagnet with high spin density and a trapped-electron platform, exemplified by the eNe platform while being equally applicable to the eHe platform, which are directly coupled through a magnetic stray field.
The quantized magnetic field produced by magnon excitations not only mediates intrinsic magnon-spin coupling but also, through its \textit{spatial gradients and curvature}, naturally introduces the electron motional states, thereby enabling higher-order multi-field interactions.
Moreover, the substantial zero-point fluctuations of the motional states in eNe platform significantly enhance higher-order interactions.
Building on these mechanisms, we demonstrate a tunable and strongly \textit{nonlinear tripartite coupling}, in which two phonons arising from the electron motional states simultaneously interact with a single spin flip and a magnon excitation---a phenomenon not previously observed in hybrid quantum systems.
To further manipulate and amplify this nonlinear tripartite interaction, parametric driving---achieved by applying a time-dependent electric field that modulates the confinement potential---is employed to enhance the zero-point fluctuations of the electron motion.
As a result, an exponential enhancement of the spin-magnon-motion coupling is achieved, with the amplification factor expressed as $e^{2r}$, where $r$ represents the strength of the parametric drive.
The proposed scheme differs from previous theoretical and experimental works, as it introduces a novel nonlinear spin-magnon-motion tripartite interaction, whose experimental implementation requires only minor modifications to existing setups.
By exploiting this nonlinear tripartite interaction, magnons mediate effective coupling among four distinct degrees of freedom across two electrons, providing a reliable platform for the preparation and investigation of few-body entanglement.

\begin{figure}
	\centering
	\includegraphics[width=0.48\textwidth]{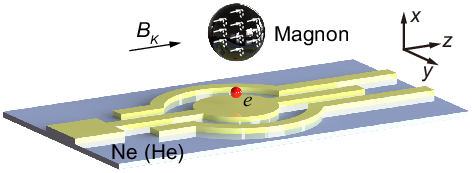}
	\caption{\textbf{The setup.} An electron is trapped above a solid neon surface, with a magnetic sphere positioned above it via a clamping structure.}
	\label{FIG1}
\end{figure}
\section{Results and Discussion}
\subsection{The setup}
As depicted in Fig.~\ref{FIG1}(a), we investigate a hybrid quantum system consisting of a YIG micromagnetic sphere coupled to a single electron trapped above solid Ne.
The microsphere can be positioned directly above the trapped electron using a mechanical clamping device.
The energy spacing between the ground and first excited states of the electron’s motion in the $x$ direction is on the order of THz, far exceeding the thermal energy at millikelvin temperatures, so this degree of freedom is effectively frozen and its dynamics can be neglected~\cite{2022ZhouP4650,2024ZhouP116122,SM}.
Confinement in the $zy$-plane is provided by an anisotropic harmonic potential, with trapping frequencies $\omega_z$ and $\omega_y$ far off-resonant, allowing the coupling to motion along the $y$ direction to be neglected, as this degree of freedom remains effectively frozen in its quantum ground state (e.g., $\omega_y \gg \omega_z$)~\cite{2022ZhouP4650,2024ZhouP116122,SM}.
We now focus on the electron’s motion along the $z$ direction, which is governed by the harmonic oscillator Hamiltonian $\hat{H}_z=\hat{p}_z^2/(2m_e)+1/2m_e\omega_z^2 \hat{z}^2$, where $m_e$ is the electron mass~\cite{SM}.
Introducing the creation and annihilation operators $\hat{a}^\dagger$ and $\hat{a}$ associated with the motional state of the trapped electron along the $z$ direction, the position operator is written as $\hat{z}=z_0(\hat{a}+\hat{a}^\dagger)$, where $z_0=\sqrt{\hbar/(2m_e \omega_z)}$ denotes the zero-point fluctuation.
In terms of these operators, the Hamiltonian for the $z$-direction motion is $\hat{H}_z=\hbar\omega_z\hat{a}^\dagger\hat{a}$.
Hereafter, we set $\hbar=1$ for simplicity.

In the trapped electron system, the electron also possesses intrinsic degrees of freedom---spin.
A static magnetic field $B_s$ is applied along the $z$-direction to define the spin quantization axis.
Thus, the Hamiltonian of the spin qubit can be expressed as $\hat{H}_s=\omega_s/2\hat{\sigma}_z$, where the resonance frequency is given by $\omega_s=\gamma_e B_s$, and the Pauli operator is defined as $\hat{\sigma}_z=\vert e\rangle\langle e\vert-\vert g\rangle\langle g\vert$.
Here, $\vert e\rangle$ and $\vert g\rangle$ represent the spin-up and spin-down states, respectively.

To enable coupling between the electron and a magnonic system, a YIG sphere is positioned directly above the trapped electron [Fig.~\ref{FIG1}(a)].
The stray field around the YIG magnetic sphere is small enough not to affect the properties of solid Ne.
The radius of the YIG sphere is denoted as $R_K$, and the distance from its surface to the trapped electron is $d_K$.
An external bias field $B_K$ saturates the magnetization of the YIG sphere, supporting long-lived spin-wave modes, among which we focus on the lowest-order Kittel mode with uniform precession across the sphere.
The spin-wave quasiparticle (magnon) is described by the free Hamiltonian $\hat{H}_K = \omega_K \hat{s}_K^\dagger \hat{s}_K$, where the resonance frequency is given by $\omega_K = \gamma_e B_K$~\cite{2020GonzalezBallesteroP125404125404,2020GonzalezBallesteroP9360293602,SM}.
Here, $\hat{s}_K$ and $\hat{s}_K^\dagger$ represent the annihilation and creation operators of the magnon, respectively.

\subsection{The coupling mechanism\label{TCM}}
\begin{figure*}
	\centering
	\includegraphics[width=0.75\textwidth]{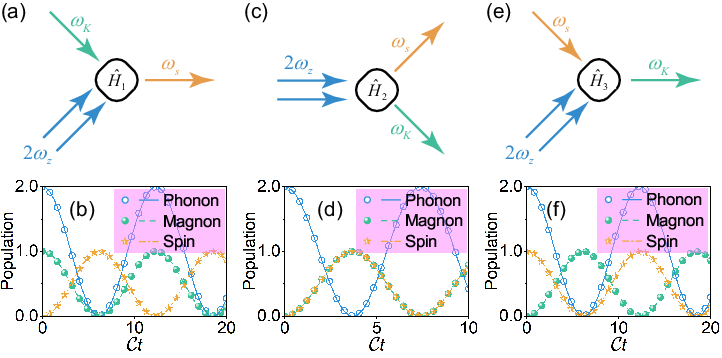}
	\caption{\textbf{System dynamics.} Panels (a, c, e) schematically depict the energy-exchange processes corresponding to the three effective Hamiltonians. (b, d, f) present the dynamical processes of the system under the resonance conditions $\Delta_1=0$, $\Delta_2=0$, and $\Delta_3=0$, respectively. The scattered points correspond to results obtained from the full Hamiltonian $\hat{H}_\mathrm{KE}$, while the solid curves correspond to results from the effective Hamiltonian $\hat{H}_{1,2,3}$. The parameter $\epsilon_0$ is assigned a value of $0.2$.}
	\label{FIG_Add_1}
\end{figure*}
A quantized stray magnetic field $\hat{\boldsymbol{B}}=(\hat{B}_x,\hat{B}_y,\hat{B}_z)$ is generated by the magnon mode in the YIG sphere~\cite{SM}.
Furthermore, the quantized stray field exhibits a gradient and curvature along the $z$ axis, thereby introducing the electron’s motion into the original two-body interaction, yielding a tripartite coupling.
For $z/H \ll 1$ ($H=d_K+R_K$), the three components of the quantized magnetic field can be approximated as $\hat{B}_x=B_c[2-6(z/H)^2 ]\hat{m}_x$, $\hat{B}_y=B_c[-1+3/2(z/H)^2]\hat{m}_y$, and $\hat{B}_z=-3B_c(z/H) \hat{m}_x$~\cite{SM}.
Here, $B_c=\mu_0 V_K M_K/(4\pi H^3)$ and $\hat{m}_{x,y,z}$ denote the three components of the quantized magnetization operator $\hat{\boldsymbol{m}}=M_K(\widetilde{\boldsymbol{m}}_K\hat{s}_K+\widetilde{\boldsymbol{m}}_K^*\hat{s}_K^\dagger)$.
Here, the YIG sphere volume, the Kittel mode function, and the zero-point magnetization are denoted by $V_K$, $\widetilde{\boldsymbol{m}}_K$, and $M_K$, respectively.

The magnon-electron interaction is characterized by the Hamiltonian $\hat{H}_\mathrm{KE}=-\gamma_e\hat{\boldsymbol{B}}\cdot\hat{\boldsymbol{S}}$, which arises from the magnetic dipole interaction.
Here, $\gamma_e$ denotes the electron gyromagnetic ratio, and the spin operator is defined as $\hat{\boldsymbol{S}}=1/2\hat{\boldsymbol{\sigma}}$, with $\hat{\boldsymbol{\sigma}}=(\hat{\sigma}_x,\hat{\sigma}_y,\hat{\sigma}_z)$ representing the Pauli matrices.
By substituting the quantized magnetic field $\hat{\boldsymbol{B}}$ into the interaction Hamiltonian $\hat{H}_\mathrm{KE}$, one obtains $\hat{H}_\mathrm{KE}=-g_\mathrm{Tx}(\hat{s}_K+\hat{s}_K^\dagger)\hat{\sigma}_x+g_\mathrm{Ty}(\hat{s}_K-\hat{s}_K^\dagger)(\hat{\sigma}_+ - \hat{\sigma}_-)+g_L(\hat{a}+\hat{a}^\dagger)(\hat{s}_K+\hat{s}_K^\dagger)\hat{\sigma}_z+g_x^{(2)}(\hat{a}+\hat{a}^\dagger)^2(\hat{s}_K+\hat{s}_K^\dagger)\hat{\sigma}_x-g_y^{(2)}(\hat{a}+\hat{a}^\dagger)^2(\hat{s}_K-\hat{s}_K^\dagger)(\hat{\sigma}_+-\hat{\sigma}_-)$
with the coupling strengths defined as $g_\mathrm{Tx}=2\mathcal{C}$, $g_\mathrm{Ty}=\mathcal{C}$, $g_L=3\mathcal{C}\epsilon_0$, $g_x^{(2)}=6\mathcal{C}\epsilon_0^2$, and $g_y^{(2)}=3\mathcal{C}\epsilon_0^2/2$.
Here, the constants are defined as $\mathcal{C}=\gamma_e B_c/2$ and $\epsilon_0=z_0/H$.
Previous studies have predominantly investigated the gradient term ($\propto z$), which mediates single-excitation energy exchange~\cite{2021HeiP4370643706,2023HeiP7360273602}.
Hereafter, we refer to this as the linear term, which is not the focus of our study.
In contrast, higher-order nonlinear interactions, arising from the curvature ($\propto z^2$) of the magnetic field and responsible for multi-excitation energy exchange, are often neglected due to their coupling strengths being typically several orders of magnitude weaker than those of the linear term, as exemplified by magnetic spheres coupled to mechanical oscillators~\cite{2023HeiP7360273602}.
For a confined electron, the zero-point fluctuation amplitude $z_0$ is significantly larger than that in typical mechanical oscillators.
This enhancement substantially amplifies the higher-order interaction terms $g_x^{(2)}$ and $g_y^{(2)}$.
By individually tuning the resonance frequencies of the subsystems to enter specific resonance regimes, a particular nonlinear process can be selectively activated.
Using the rotating-wave approximation and neglecting far-off-resonant terms, the interaction Hamiltonian is simplified into different nonlinear tripartite coupling~\cite{SM}:
\begin{subequations}
	\begin{align}
		\Delta_1&=0,~\hat{H}_1=\mathcal{A}\left(\hat{a}^2\hat{s}_K\hat{\sigma}_++\hat{a}^{\dagger 2}\hat{s}_K^\dagger\hat{\sigma}_-\right), \label{HNLone}\\
		\Delta_2&=0,~\hat{H}_2=\mathcal{B}\left(\hat{a}^2\hat{s}_K^\dagger\hat{\sigma}_++\hat{a}^{\dagger 2}\hat{s}_K\hat{\sigma}_-\right), \label{HNLTwo}\\
		\Delta_3&=0,~\hat{H}_3=\mathcal{A}\left(\hat{a}^2\hat{s}_K^\dagger\hat{\sigma}_-+\hat{a}^{\dagger 2}\hat{s}_K\hat{\sigma}_+\right), \label{HNLthree}
	\end{align}
	\label{HNL}
\end{subequations}
where the corresponding resonance conditions are $\Delta_1=2\omega_z+\omega_K-\omega_s$, $\Delta_2=2\omega_z-\omega_K-\omega_s$, and $\Delta_3=2\omega_z-\omega_K+\omega_s$, with $\omega_K \neq \omega_s$.
Here the coupling strengths are redefined as $\mathcal{A}=9\mathcal{C}\epsilon_0^2/2$ and $\mathcal{B}=15\mathcal{C}\epsilon_0^2/2$.

To verify the validity of the rotating-wave approximation, we performed numerical simulations using both the original Hamiltonian $\hat{H}_\mathrm{KE}$ and the effective Hamiltonians $\hat{H}_{1,2,3}$.
As shown in Fig.~\ref{FIG_Add_1}, scattered points denote results obtained from $\hat{H}_\mathrm{KE}$, while the curves represent those from the effective Hamiltonians $\hat{H}_{1,2,3}$, exhibiting excellent agreement and confirming the validity of the rotating-wave approximation under the chosen resonance conditions.
Under the resonance condition $\Delta_1$, the energy-exchange process depicted in Fig.~\ref{FIG_Add_1}(a) involves the annihilation of two phonons and one magnon, resulting in a single spin excitation; in the reverse process, a spin excitation is annihilated to produce two phonons and one magnon.
The dynamics of this process are depicted in Fig.~\ref{FIG_Add_1}(b).
Under resonance condition $\Delta_2$, energy flows from two phonons to a spin excitation and a magnon, whereas the reverse process converts a spin excitation and a magnon back into two phonons, as illustrated in Fig.~\ref{FIG_Add_1}(c).
The numerical simulations of this energy-exchange process are depicted in Fig.~\ref{FIG_Add_1}(d).
Figure~\ref{FIG_Add_1}(e) illustrates that, under resonance condition $\Delta_3$, two phonons and a spin excitation are converted into a magnon excitation, while the reverse process produces two phonons and a spin excitation from a magnon excitation.
The dynamical behavior of this process is illustrated in Fig.~\ref{FIG_Add_1}(f).
These results indicate that energy exchange among magnons, phonons, and spin excitations conserves the total number of excitations.

\subsection{The coupling strength}
\begin{figure*}
	\centering
	\includegraphics[width=1\textwidth]{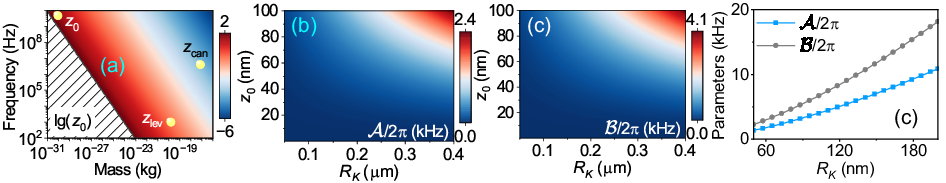}
	\caption{\textbf{Coupling strength analysis.} (a) The zero-point fluctuation varies with the mass and oscillation frequency of the mechanical resonator. The hatched region indicates parameter regimes that are not of interest in this work. Here, we focus on the nonlinear coupling strength, which is proportional to the square of the zero-point fluctuation, $z_0^2$. Panels (b) and (c) show, respectively, the dependence of the nonlinear tripartite coupling strengths $\mathcal{A}$ and $\mathcal{B}$ on the zero-point fluctuation $z_0$ and the radius $R_K$ of the YIG magnetic sphere, with the sphere-electron distance fixed at $H=500~\mathrm{nm}$. (d) depicts the nonlinear tripartite coupling strength as a function of the magnetic sphere radius $R_K$ in the confined-electron system, with other parameters fixed at $z_0=42.9~\mathrm{nm}$ and $\epsilon_0=0.2$.}
	\label{FIG2}
\end{figure*}

This section systematically investigates the determinants of the nonlinear coupling strength and clarifies the physical mechanism of the enhanced higher-order nonlinear interactions in trapped-electron systems.
As shown in Sec.~\ref{TCM}, the nonlinear coupling strength scales as $\epsilon_0^2\propto z_0^2$, such that enhanced zero-point fluctuations directly translate into stronger coupling.
Figure~\ref{FIG2}(a) shows that the zero-point fluctuation decreases with increasing mass and oscillation frequency of the mechanical oscillator (on a logarithmic scale).
This behavior reveals the mechanism underlying the enhanced higher-order nonlinear coupling in the trapped-electron platform.
In Fig.~\ref{FIG2}(a), we compare the zero-point fluctuations of trapped electrons with those of two representative classes of classical mechanical oscillators, namely, levitated particles and cantilever resonators.
The detailed calculation of the zero-point fluctuations for levitated particles and cantilever resonators is provided in the Methods (Sec.~\ref{TEvsCM}).
As shown in Fig.~\ref{FIG2}(a), the zero-point fluctuations of trapped electrons are significantly larger than those of conventional mechanical oscillators.
Consequently, higher-order nonlinear interactions, which are typically negligible in hybrid quantum systems based on conventional mechanical oscillators, become significantly enhanced in trapped-electron systems.
This constitutes the central mechanism underlying the enhancement of the nonlinear coupling strength in the present model.

We next numerically evaluate the coupling strength associated with the nonlinear tripartite interaction.
We consider a YIG magnetic sphere with a saturation magnetization $M_s=587~\mathrm{kA/m}$~\cite{2023HeiP7360273602,2021HeiP4370643706,2020GonzalezBallesteroP125404125404,2020GonzalezBallesteroP9360293602}.
The trapped electron oscillates at a frequency of $\omega_z/2\pi=5~\mathrm{GHz}$, corresponding to a zero-point fluctuation of $z_0=42.9~\mathrm{nm}$.
In Figs.~\ref{FIG2}(b) and \ref{FIG2}(c), the separation between the magnetic sphere and the electron is fixed at $H=500~\mathrm{nm}$, illustrating the dependence of the nonlinear coupling strengths $\mathcal{A}$ and $\mathcal{B}$ on the zero-point fluctuation $z_0$ and the sphere radius $R_K$, respectively.
It is evident that the coupling strength increases with increasing zero-point fluctuation.
In Fig.~\ref{FIG2}(d), with the separation fixed at $H=215~\mathrm{nm}$, the coupling strength is observed to increase with the magnetic sphere radius $R_K$.
The results demonstrate that the nonlinear coupling strength can reach the kilohertz regime, a magnitude that is unattainable in hybrid quantum systems based on conventional mechanical platforms.
As an example, Ref.~\cite{2023HeiP7360273602} studies a hybrid system consisting of NV centers in diamond coupled to a YIG magnetic sphere.
Under identical parameter settings, the zero-point fluctuation in the NV-center system is approximately $z_\mathrm{lev}=0.75~\mathrm{nm}$, substantially smaller than the $z_0=42.9~\mathrm{nm}$ in the trapped-electron system.
The corresponding nonlinear coupling strength is merely $5.82~\mathrm{Hz}$ (see Sec.\ref{CCS} for details of the calculation).
Furthermore, in levitated diamond systems, increasing the oscillation frequency to access higher-frequency regimes further suppresses the zero-point fluctuations.
This reduction in zero-point fluctuations weakens the higher-order coupling strength, thereby limiting the applicability of such schemes across a wide frequency range.
In contrast, the large zero-point fluctuations of the motional state in trapped-electron systems significantly enhance the nonlinear coupling strength.

\subsection{Exponentially enhanced coupling strength}
Next, we employ parametric amplification to enhance the nonlinear coupling strength.
By applying a time-dependent modulation to the trapped voltage $V(t)=V_c-V_t\cos (2\omega_p t)$, the Hamiltonian of the motional state is given by $\hat{H}_z^\mathrm{NL}=\hat{H}_z-1/2k_t\cos(2\omega_p t)\hat{z}^2$~\cite{SM}.
By quantizing the motion of the trapped electron along the $z$-direction, the Hamiltonian can be written as
\begin{equation}
\hat{H}_z^\mathrm{NL}=\omega_z\hat{a}^\dagger\hat{a}-\Omega_p\cos\left(2\omega_p t\right)\left(\hat{a}+\hat{a}^\dagger\right)^2
\end{equation}
with a driving strength of $\Omega_p=k_tz_0^2/2$.
We now illustrate how an exponential enhancement of the coupling strength can be achieved, taking the physical processes described by the Hamiltonian $\hat{H}_2$ as an example.
Transforming the Hamiltonian into the rotating frame with frequency $\omega_p$, we obtain
$\hat{H}_\mathrm{NKE}=\Delta_z\hat{a}^\dagger\hat{a}+\mathcal{W}_K\hat{s}_K^\dagger\hat{s}_K+\mathcal{W}_s/2\hat{\sigma}_z+\mathcal{B}(\hat{a}^2\hat{s}_K^\dagger\hat{\sigma}_+ + \hat{a}^{\dagger 2}\hat{s}_K \hat{\sigma}_-)-\Omega_p/2(\hat{a}^2+\hat{a}^{\dagger 2})$.
Here, the detunings are defined as $\Delta_z=\omega_z-\omega_p$ and $\mathcal{W}_{K,s}=\omega_{K,s}-\omega_p$.
By applying the Bogoliubov transformation $\hat{b}=\hat{a}\cosh r-\hat{a}^\dagger\sinh r$~\cite{2016LemondeP1133811338,2019BurdP11631165,2021BlaisP2500525005,2021BurdP898902}, the effective Hamiltonian of the system can then be expressed as
\begin{equation}
	\begin{split}
		\hat{H}_\mathrm{NKE}^\mathrm{eff}&=\mathcal{W}_b\hat{b}^\dagger\hat{b}+\mathcal{W}_K\hat{s}_K^\dagger\hat{s}_K+\frac{\mathcal{W}_s}{2}\hat{\sigma}_z\\ &+\widetilde{\mathcal{B}}\left(\hat{b}^2\hat{s}_K^\dagger\hat{\sigma}_+ + \hat{b}^{\dagger 2}\hat{s}_K\hat{\sigma}_-\right),
	\end{split}
	\label{H_NKE}
\end{equation}
where the effective frequency is $\mathcal{W}_b=\Delta_z/\cosh(2r)$, the effective coupling strength is $\widetilde{\mathcal{B}}=\mathcal{B}\cosh^2 r$, and the squeezing parameter satisfies $\tanh 2r=\Omega_p/\Delta_z$.
Here, the resonance condition satisfied by the Hamiltonian~(\ref{H_NKE}) is $\widetilde{\Delta}_2=2\mathcal{W}_b-\mathcal{W}_K-\mathcal{W}_s=0$.
An exponential enhancement of the coupling strength, $e^{2r}$, arising from the nonlinear term in the tripartite interaction, is enabled, resulting in an increase of two to four orders of magnitude compared with the unmodulated case.

\begin{figure}
	\centering
	\includegraphics[width=0.48\textwidth]{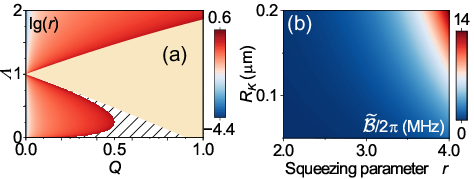}
	\caption{\textbf{Enhancement of the coupling strength.} (a) Distribution of the squeezing parameter $r$ within the stable regions. Here, $\Lambda=\omega_z^2/\omega_p^2$ and $Q=k_t/(2m_e\omega_p^2)$~\cite{SM}. (b) shows the coupling strength $\widetilde{\mathcal{B}}$ versus the squeezing parameter $r$ and the radius of the magnetic sphere $R_K$ with the other parameters: $z_0=42.9~\mathrm{nm}$ and $\epsilon_0=0.2$.}
	\label{FIG3}
\end{figure}

Figure~\ref{FIG3}(a) illustrates the accessible range of the squeezing parameter in the presence of two-phonon driving.
In Fig.~\ref{FIG3}(a), the yellow-shaded region indicates the unstable regime, where the electron cannot be stably confined with two-phonon driving.
The hatched area corresponds to nonphysical parameter values, where the parameter $\tanh(2r)$ exceeds unity.
For parametric amplification, precise control of the ratio $\Omega_p/\Delta_z$ is required.
As this ratio approaches unity, the attainable squeezing parameter $r$ increases.
As illustrated in Fig.~\ref{FIG3}(a), when $\Omega_p/\Delta_z$ approaches unity, the squeezing parameter $r$ increases, and the system approaches the boundary between the stable and unstable regimes.
Figure~\ref{FIG3}(b) shows that, under parametric amplification, the nonlinear coupling strength can reach the megahertz scale.
Importantly, while parametric amplification enhances the nonlinear coupling strength, it also increases the dissipation of the electron’s motional states.
Let $\gamma_\mathrm{ph}$ represent the intrinsic dissipation of the motional states of the electrons, and $\gamma_\mathrm{ph}^\mathrm{amp}=\gamma_\mathrm{ph}\exp(2r)$ the corresponding dissipation after amplification.
In the following discussion, unless stated otherwise, we will employ the amplified dissipation $\gamma_\mathrm{ph}^\mathrm{amp}$ of the motional states of the electrons.

\subsection{Magnon-mediated electron-electron interaction}
\begin{figure}
	\centering
	\includegraphics[width=0.5\textwidth]{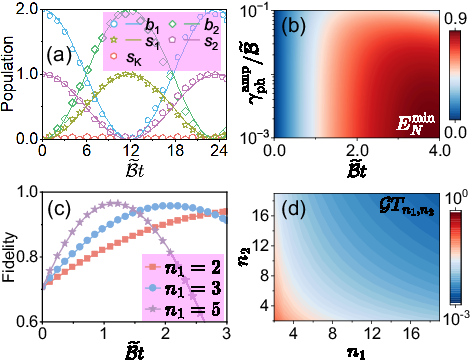}
	\caption{\textbf{Mediated (indirect) coupling.} (a) Dynamics of the original Hamiltonian $\hat{H}_{\mathrm{EKE}}$ versus the effective Hamiltonian $\hat{H}_{\mathrm{EE}}$. Here, $b_i$ and $s_i$ denote the phonon and spin degrees of freedom of the $i$-th electron, respectively, while $s_K$ represents the Kittel mode. The parameters are set as $\mathcal{D}_K=15\widetilde{\mathcal{B}}$, $\mathcal{W}_s=10\widetilde{\mathcal{B}}$, and $\mathcal{D}_b=10\widetilde{\mathcal{B}}^2/(2\mathcal{D}_K)$. Panel (b) shows the effect of dissipation on entanglement. (c) Dependence of the fidelity on the initial excitation number $n_1$. The parameters are set as follows: the coupling strength $\mathcal{G}_\mathrm{EE}=0.1~\widetilde{\mathcal{B}}$, the phonon decay rate $\gamma_\mathrm{ph}^\mathrm{amp}=0.01~\widetilde{\mathcal{B}}$, the spin dephasing rate $\gamma_s=0.01~\widetilde{\mathcal{B}}$, and initial excitation number $n_2=0$. (d) Evolution time $T_{n_1,n_2}$ required to reach the entangled state $\psi_E$ as a function of the initial excitation numbers $n_1$ and $n_2$.}
	\label{FIG4}
\end{figure}

We extend our analysis to a system consisting of two electrons confined in different electrodes and simultaneously coupled to a single YIG sphere.
After the parametric amplification, the system's effective Hamiltonian is expressed as $\hat{H}_{\mathrm{EKE}}=\mathcal{W}_b\sum_{i=1,2}\hat{b}_i^\dagger\hat{b}_i+\mathcal{W}_K\hat{s}_K^\dagger\hat{s}_K+\mathcal{W}_s/2\sum_{i=1,2}\hat{\sigma}_z^{(i)}+\widetilde{\mathcal{B}}\sum_{i=1,2}[\hat{b}_i^2\hat{s}_K^\dagger\hat{\sigma}_+^{(i)}+(\hat{b}_i^\dagger)^2\hat{s}_K\hat{\sigma}_-^{(i)}]$.
Here, $i$ labels the $i$-th electron.
Transforming the Hamiltonian into a rotating frame with frequency $\mathcal{W}_s$ gives
\begin{equation}
	\begin{split}
		\hat{H}_{\mathrm{EKE}}&=\mathcal{D}_b\left(\hat{b}_1^\dagger\hat{b}_1+\hat{b}_2^\dagger\hat{b}_2\right)+\mathcal{D}_K\hat{s}_K^\dagger\hat{s}_K\\
		&+\widetilde{\mathcal{B}}\sum_{i=1,2}\left[\hat{b}_i^2\hat{s}_K^\dagger\hat{\sigma}_+^{(i)}+(\hat{b}_i^\dagger)^2\hat{s}_K\hat{\sigma}_-^{(i)}\right],
	\end{split}
\end{equation}
where $\mathcal{D}_{b,K}=\mathcal{W}_{b,K}-\mathcal{W}_s$.
In the far-detuned regime for the magnon (i.e., $\mathcal{D}_K \gg 2\mathcal{D}_b,~ \widetilde{\mathcal{B}}$), the magnon mode can be adiabatically eliminated, resulting in an effective system Hamiltonian~\cite{SM}
\begin{equation}
	\hat{H}_\mathrm{EE}=-\mathcal{G}_\mathrm{EE}\left[\left(\hat{b}_1^\dagger\right)^2\hat{b}_2^2\hat{\sigma}_-^{(1)}\hat{\sigma}_+^{(2)}+\hat{b}_1^2\left(\hat{b}_2^\dagger\right)^2\hat{\sigma}_+^{(1)}\hat{\sigma}_-^{(2)}\right].
\end{equation}

To validate the effective Hamiltonian, Fig.~\ref{FIG4}(a) compares numerical simulations of the original Hamiltonian $\hat{H}_{\mathrm{EKE}}$ (scatter points) with those of the effective Hamiltonian $\hat{H}_\mathrm{EE}$ (curve), both initialized in the state $\psi_0=\vert n_1=2,n_2=0,g,e\rangle$.
A close agreement between the two confirms the validity of the effective model.
This Hamiltonian encompasses four modes, corresponding to the phonon and spin degrees of freedom of the two electrons, making it particularly suitable for investigating few-body entanglement.
When the system is initially prepared in the quantum state $\psi_0$, its state at an arbitrary time $t$ can be expressed as $\psi(t)=\cos(\Omega_{n_1,n_2}t) \vert n_1,n_2,g,e\rangle+i\sin (\Omega_{n_1,n_2}t) \vert n_1-2,n_2+2,e,g\rangle$, where $\Omega_{n_1,n_2}=\mathcal{G}_\mathrm{EE}\sqrt{n_1(n_1-1)(n_2+1)(n_2+2)}$.
The system evolves into the entangled state $\psi_E=(\vert n_1,n_2,g,e\rangle+i\vert n_1-2,n_2+2,e,g\rangle)/\sqrt{2}$ at $T_{n_1,n_2}=\pi/(4\Omega_{n_1,n_2})$, which decreases with increasing initial excitations $n_{1,2}$.
As shown in Fig.~\ref{FIG4}(b), in the absence of dissipation or under weak dissipation, $E_N^\mathrm{min}$ (See Sec.~\ref{ENmin} for details of the calculation of $E_N^\mathrm{min}$) can reach a relatively large value (e.g., close to 1), indicating strong entanglement in the system.
As the dissipation strength increases, the peak value of $E_N^\mathrm{min}$ gradually decreases, indicating that the maximal achievable entanglement of the system is degraded.
Figures~\ref{FIG4}(c) and \ref{FIG4}(d) illustrate that the time required for the system to reach the entangled state $\psi_E$ decreases as the excitation numbers $n_1$ and $n_2$ increase, with the decrease becoming more pronounced when both excitations are higher.
In summary, magnon-mediated electron-electron interactions enable the rapid preparation of few-body entangled states.

\subsection{Experimental feasibility}
In this section, we analyze the experimentally feasible parameters of the system.
The motional resonance frequency of the electron is set to $\omega_z/2\pi=5~\mathrm{GHz}$, corresponding to a zero-point fluctuation $z_0=42.9~\mathrm{nm}$.
A YIG magnetic sphere is positioned above the electron using a clamping setup, with a fixed separation $H=215~\mathrm{nm}$ and a radius of $150~\mathrm{nm}$.
With these parameters, the bare nonlinear coupling strength is estimated to be $\mathcal{B}/2\pi=12~\mathrm{kHz}$.
The system enters the strong-coupling regime when the condition $\mathcal{B}>\gamma_\mathrm{ph},~\gamma_s,~\gamma_K$ is satisfied, where $\gamma_\mathrm{ph}$, $\gamma_s$, and $\gamma_K$ denote the dissipation of the electron’s motional state, the dephasing of the electron spin, and the magnon's dissipation, respectively.
Here, The relevant dissipation rates are $\gamma_\mathrm{ph}/2\pi=16.7~\mathrm{Hz}$ for the electron’s motional mode~\cite{2025LiP250720476} and $\gamma_K/2\pi=0.49~\mathrm{MHz}$ for the magnon mode~\cite{2025ShenP56525652}.
The electron spin dephasing rate can reach approximately $\gamma_s/2\pi \approx 0.01~\mathrm{Hz}$ through the application of Hahn echoes~\cite{2022ChenP4501645016}.
Even in the presence of an external magnetic field, a relatively conservative dephasing rate of $\gamma_s/2\pi=10~\mathrm{kHz}$ is adopted~\cite{2010SchusterP4050340503}.
These results indicate that the bare coupling strength is smaller than the magnon dissipation and is therefore insufficient to reach the strong-coupling regime.
Parametric amplification enables an exponential enhancement of the coupling strength, with an enhancement factor of $e^{2r}$.
The squeezing parameter $r$ is defined by $\tanh 2r=\Omega_p/\Delta_z$, and increases as the ratio $\Omega_p/\Delta_z$ approaches unity.
Figure~\ref{FIG3}(a) illustrates the accessible parameter regime for $r$.
Notably, parametric amplification also enhances the noise; therefore, the amplified dissipation $\gamma_\mathrm{ph}^\mathrm{amp}$ is adopted in the following analysis.
For a squeezing parameter $r=2.6$, the nonlinear coupling strength $\widetilde{\mathcal{B}}/2\pi=0.54~\mathrm{MHz}$, exceeds the amplified dissipation $\gamma_\mathrm{ph}^\mathrm{amp}/2\pi=3.02~\mathrm{kHz}$ and other relevant decay rates
($\gamma_s,~\gamma_K$), enabling the system to enter the strong-coupling regime.
Increasing $r$ to $3.6$ further enhances the coupling to $4~\mathrm{MHz}$ (with $\gamma_\mathrm{ph}^\mathrm{amp}/2\pi=22.3~\mathrm{kHz}$), confirming robust strong coupling.

\section{Conclusion}
We propose a hybrid quantum system composed of a micromagnetic sphere and an eNe platform.
This platform enables the realization of multiple forms of nonlinear interactions under experimentally accessible conditions.
By employing parametric amplification, the coupling strength can be exponentially enhanced.
Furthermore, with magnons acting as intermediaries, nonlinear interactions between distinct electrons are mediated, resulting in strong indirect coupling among the four degrees of freedom associated with two electrons.
These interactions enable the rapid generation of few-body entangled states.
This hybrid system provides a novel platform for exploring quantum effects through engineered hybrid interactions.

\section{Methods}
\subsection{Trapped electrons vs. conventional mechanical systems\label{TEvsCM}}
This section presents a comparison between two conventional mechanical platforms and the trapped-electron system.
For the levitated particle, we adopt the model described in Ref.~\cite{2023HeiP7360273602}.
The system consists of a trapped diamond particle with a resonance frequency $\omega_\mathrm{lev}/2\pi=1~\mathrm{kHz}$, a particle radius of $R_\mathrm{lev}=10~\mathrm{nm}$, and a diamond density of $\rho_\mathrm{lev}=3520~\mathrm{kg/m^3}$.
Using these parameters, the zero-point fluctuation is calculated to be $z_\mathrm{lev}=0.75~\mathrm{nm}$.

For the cantilever resonator, we consider a silicon cantilever with dimensions $(l_\mathrm{can},w_\mathrm{can},t_\mathrm{can})=(6,0.1,0.01)~\mathrm{\mu m}$ for length, width, and thickness, respectively.
The cantilever’s resonance frequency is given by $\omega_\mathrm{can}/2\pi=3.516\times t_\mathrm{can}/l_\mathrm{can}^2\sqrt{E/(12\rho_\mathrm{can})}$, and its effective mass is $m_\mathrm{can}=\rho_\mathrm{can}l_\mathrm{can}w_\mathrm{can}t_\mathrm{can}/4$, where the mass density is taken as $\rho_\mathrm{can}=2330~\mathrm{kg/^3}$ and the Young’s modulus is $E=1.3\times 10^{11}~\mathrm{Pa}$.
The corresponding zero-point fluctuation is calculated to be $z_\mathrm{can}=0.53~\mathrm{fm}$.
A summary of the detailed parameters is provided in Table~\ref{ZPF}.

\begin{table}[b]
	\caption{\label{ZPF}
		\textbf{Comparison of zero-point fluctuations in different mechanical systems.}
	}
	\begin{ruledtabular}
		\begin{tabular}{cccc}
			\textrm{Type}&
			\textrm{Frequency}&
			\textrm{Mass}&
			\textrm{\makecell{zero-point\\fluctuation}}\\
			\colrule
			Trapped electron & $5~\mathrm{GHz}$ & $9.1\times 10^{-31}~\mathrm{kg}$ & $42.9~\mathrm{nm}$\\
			Levitated particle & $1~\mathrm{kHz}$ & $1.5\times 10^{-20}~\mathrm{kg}$ & $0.75~\mathrm{nm}$\\
			Mechanical cantilever & $4~\mathrm{MHz}$ & $7.0\times 10^{-18}~\mathrm{kg}$ & $0.53~\mathrm{fm}$\\
		\end{tabular}
	\end{ruledtabular}
\end{table}

\subsection{Comparison of coupling strengths\label{CCS}}
This section presents a comparison of the nonlinear coupling strength between our system and the system reported in Ref.~\cite{2023HeiP7360273602}.
In Ref.~\cite{2023HeiP7360273602}, the authors investigated a hybrid system consisting of nitrogen-vacancy (NV) centers in diamond coupled to a YIG magnetic sphere.
The interaction Hamiltonian arises from the magnetic dipole coupling between the NV center spin and the magnon mode of the YIG magnetic sphere.
Following the model in Ref.~\cite{2023HeiP7360273602}, the second-order nonlinear coupling induced by the curvature of the magnetic field is given by
\begin{equation}
	G_\mathrm{NV}=\frac{2\gamma_e\mu_0 M_K R_K^3}{r_0^5}z_\mathrm{lev}^2.
\end{equation}
Here, $r_0=R_K+d_\mathrm{NV}+R_\mathrm{lev}$ represents the center-to-center distance between the magnetic sphere and the NV center.
$\mu_0$ is the vacuum permeability, $M_K$ denotes the zero-point magnetization fluctuation, $\gamma_e$ is the electron gyromagnetic ratio, and $z_\mathrm{lev}$ corresponds to the zero-point fluctuation of the levitated diamond particle.
Here, $d_\mathrm{NV}$ denotes the surface-to-surface distance between the YIG magnetic sphere and the levitated diamond particle.
For a quantitative comparison, we consider a magnetic sphere with radius $R_K=150~\mathrm{nm}$, and the diamond particle is characterized by the following typical parameters: density $\rho_\mathrm{lev}=3520~\mathrm{kg/m^3}$, radius $R_\mathrm{lev}=10~\mathrm{nm}$, surface separation $d_\mathrm{NV}=55~\mathrm{nm}$, and center-of-mass oscillation frequency $\omega_\mathrm{lev}/2\pi=1~\mathrm{kHz}$.
Under these parameters, the zero-point fluctuation of the diamond particle is approximately $z_\mathrm{lev}=0.75~\mathrm{nm}$, much smaller than that of the trapped-electron system in this work ($42.9~\mathrm{nm}$).
Consequently, the corresponding nonlinear coupling strength is merely $G_\mathrm{NV}/2\pi=5.82~\mathrm{Hz}$, approximately three orders of magnitude weaker than that realized in our system.

\subsection{Numerical Simulation Methods}
All dynamical simulations presented in this work are performed using the QuTiP package~\cite{Lambert2026P162}.
In Figs.~\ref{FIG_Add_1}(b), \ref{FIG_Add_1}(d), and \ref{FIG_Add_1}(f), the system dynamics are obtained by numerically solving the Schr\"odinger equation
\begin{equation}
	i\frac{\partial}{\partial t}\psi(t)=\hat{H}_{\mathrm{sys}}\psi(t).
	\label{SchEq}
\end{equation}
The initial states for Figs.~\ref{FIG_Add_1}(b), \ref{FIG_Add_1}(d), and \ref{FIG_Add_1}(f) are chosen as $\vert 2\rangle_\mathrm{ph}\vert 1\rangle_M\vert g\rangle_S$, $\vert 2\rangle_\mathrm{ph}\vert 0\rangle_M\vert g\rangle_S$, and $\vert 2\rangle_\mathrm{ph}\vert 0\rangle_M\vert e\rangle_S$, respectively.
Here, the subscripts $\mathrm{ph}$, $M$, and $S$ denote the phonon, magnon, and spin degrees of freedom, respectively.
The discrete markers correspond to numerical results obtained from the full Hamiltonian $\hat{H}_\mathrm{sys}=\hat{H}_\mathrm{KE}$, while the curves represent the dynamical evolution governed by the effective Hamiltonian $\hat{H}_\mathrm{sys}=\hat{H}_{1,2,3}$.

In Fig.~\ref{FIG4}(a), the system dynamics are likewise obtained by solving the Schrödinger equation [Eq.~(\ref{SchEq})], where the discrete markers are computed using $\hat{H}_\mathrm{sys}=\hat{H}_\mathrm{EKE}$, while the curves are obtained from $\hat{H}_\mathrm{sys}=\hat{H}_\mathrm{EE}$.
The initial state is taken as $\psi_0=\vert n_1=2,n_2=0,g,e\rangle$.

The results in Figs.~\ref{FIG4}(b) and \ref{FIG4}(c) are obtained by solving the Lindblad master equation, given by
\begin{equation}
	\dot{\hat{\rho}}=i\left[\hat{\rho},\hat{H}_\mathrm{EE}\right]+\gamma_\mathrm{ph}^\mathrm{amp}\sum_{i=1,2}D\left[\hat{b}_i\right]\hat{\rho}+\gamma_s\sum_{i=1,2}D\left[\hat{s}_z^i\right]\hat{\rho}.
\end{equation}
Here, $\hat{\rho}$ denotes the density matrix of the system, and $D[\hat{O}]\cdot=\hat{O}\cdot\hat{O}^\dagger-\{\hat{O}^\dagger\hat{O},\cdot\}/2$ represents the Lindblad dissipator.

\subsection{Calculation of $E_N^\mathrm{min}$\label{ENmin}}
For the entangled state $\psi_E$, the entanglement can be quantified using the logarithmic negativity~\cite{2025MaP24892499,2002VidalP3231432314}.
The logarithmic negativity is defined as $E_N(\rho)=\log_2\vert\vert\rho^{T_A}\vert\vert_1$, where $\rho^{T_A}$ denotes the partial transpose of the density matrix $\rho$ with respect to subsystem A under a given bipartition. $\vert\vert\cdot \vert\vert_1$ denotes the trace norm.
The system comprises four degrees of freedom, $\{b_1,~b_2,~s_1,~s_2\}$. We consider all possible 1 vs 3 and 2 vs 2 bipartitions, as summarized in the Table~\ref{BP}.
To assess the presence of entanglement, we consider the minimum logarithmic negativity over all the above bipartitions
\begin{equation}
\begin{aligned}
	E_N^\mathrm{min} = \min \big[
	& E_N^{(b_1)},~E_N^{(b_2)},~E_N^{(s_1)},~E_N^{(s_2)}, \\
	& E_N^{(b_1,b_2)},~E_N^{(b_1,s_1)},~E_N^{(b_1,s_2)}
	\big].
\end{aligned}
\end{equation}

\begin{table}[b]
	\caption{\label{BP}
		\textbf{Logarithmic negativity under different bipartitions of the system.}
	}
	\begin{ruledtabular}
		\begin{tabular}{cccccc}
			\textrm{1 vs 3}&
			\textrm{$\rho^{T_A}$}&
			\textrm{$E_N(\rho)$}&
			\textrm{2 vs 2}&
			\textrm{$\rho^{T_A}$}&
			\textrm{$E_N(\rho)$}\\
			\colrule
			$\{b_1\}\vert\{b_2,s_1,s_2\}$ & $\rho^{T_{b_1}}$ & $E_N^{(b_1)}$ & $\{b_1,b_2\}\vert\{s_1,s_2\}$&$\rho^{T_{b_1,b_2}}$& $E_N^{(b_1,b_2)}$\\
			$\{b_2\}\vert\{b_1,s_1,s_2\}$ & $\rho^{T_{b_2}}$ & $E_N^{(b_2)}$ & $\{b_1,s_1\}\vert\{b_2,s_2\}$&$\rho^{T_{b_1,s_1}}$& $E_N^{(b_1,s_1)}$\\
			$\{s_1\}\vert\{b_1,b_2,s_2\}$ & $\rho^{T_{s_1}}$ & $E_N^{(s_1)}$ & $\{b_1,s_2\}\vert\{b_2,s_1\}$&$\rho^{T_{b_1,s_2}}$& $E_N^{(b_1,s_2)}$\\
			$\{s_2\}\vert\{b_1,b_2,s_1\}$ & $\rho^{T_{s_2}}$ & $E_N^{(s_2)}$ & & &
		\end{tabular}
	\end{ruledtabular}
\end{table}

\begin{acknowledgments}
	X.-F. P. is supported by the National Natural Science Foundation of China under Grants No. 124B2091.
	P.-B. L. is supported by the National Natural Science Foundation of China under Grants No. W2411002 and No. 12375018.
\end{acknowledgments}

%

\end{document}


\title{Supplemental Material for ``Nonlinear Tripartite Coupling of Trapped Electrons with Magnons in a Hybrid Quantum System''}
\author{Xue-Feng Pan}
\author{Peng-Bo Li}
\email{lipengbo@mail.xjtu.edu.cn}
\affiliation{Ministry of Education Key Laboratory for Nonequilibrium Synthesis and Modulation of Condensed Matter, Shaanxi Province Key Laboratory of Quantum Information and Quantum Optoelectronic Devices, School of Physics, Xi'an Jiaotong University, Xi'an 710049, China}

\date{\today}

\begin{abstract}
	This supplementary material presents the calculations that are omitted from the main text.
	In Sec.~\ref{SW}, the quantization of spin waves in micromagnetic spheres is discussed in detail.
	Section~\ref{eNe} provides a detailed discussion of the Rydberg and orbital states within the trapped electronic platform.
	Section~\ref{CM} presents a detailed description of the coupling mechanism between the magnon and the trapped electron platform.
	Section~\ref{EECS} provides the computational details for achieving exponential enhancement of the system's coupling strength through the parametric amplification technique.
	In Sections~\ref{MEEI} and~\ref{NGSP}, two distinct magnon-mediated coupling mechanisms were demonstrated: one facilitating coherent coupling between two electrons, and the other enabling dissipative coupling between a single electron's internal motional and spin states.
	These mechanisms serve distinct quantum state preparations: the coherent coupling of Section~\ref{MEEI} allows for the generation of few-body entangled states, whereas the dissipative coupling of Section~\ref{NGSP} enables the preparation of non-Gaussian steady states of the electron's motional state.
\end{abstract}
\maketitle

\tableofcontents
\newpage

\section{\label{SW}Spin-wave quantization}
Spin waves are collective excitations of magnetic moments in magnetically ordered materials, described by the nonlinear Landau-Lifshitz (LL) equation.
It is very difficult to solve the nonlinear LL equations.
However, with appropriate physical approximations, the Landau-Lifshitz equations can be linearized, enabling the computation of spin-wave eigenmodes.
In this section, we quantize the spin wave based on the linearized LL equation and compute the relevant parameters of the Kittel mode.
For convenience, in the following calculations, we adopt a local coordinate system with the origin at the center of the YIG magnetic sphere.

\subsection{The spin wave equations}
First we derive the equations of motion satisfied by spin waves in a spherical micromagnet.
In magnetic materials, the magnetization distribution is described by a continuous magnetization field $\boldsymbol{M}(\boldsymbol{r},t)$.
The electric and magnetic fields associated with it are $\boldsymbol{E}(\boldsymbol{r},t)$ and $\boldsymbol{B}(\boldsymbol{r},t)$, respectively, and satisfy Maxwell's equations.
The magnetization field $\boldsymbol{M}(\boldsymbol{r},t)$ obeys the nonlinear LL equation~\cite{2000Aharonip,2009Stancilp}
\begin{equation}
	\partial_t\boldsymbol{M}(\boldsymbol{r},t)=-\vert \gamma_e\vert \mu_0 \boldsymbol{M}(\boldsymbol{r},t) \times \boldsymbol{H}_{\rm{eff}}(\boldsymbol{M},\boldsymbol{r},t),
	\label{LLO}
\end{equation}
which involves the physical constants: the gyromagnetic ratio $\gamma_e$ and the vacuum permeability $\mu_0$.
The effective field $\boldsymbol{H}_\mathrm{eff}=\boldsymbol{H}(\boldsymbol{r},t)+\boldsymbol{\mathcal{H}}(\boldsymbol{M}, \boldsymbol{r},t)$ consists of two components: the first is Maxwell's field $\boldsymbol{H}(\boldsymbol{r},t)$, and the second is the effective field $\boldsymbol{\mathcal{H}}(\boldsymbol{M}, \boldsymbol{r},t)$ arising from various types of interactions in magnetic materials.
The effective field $\boldsymbol{\mathcal{H}}(\boldsymbol{M}, \boldsymbol{r},t)$ can be written as~\cite{2009Stancilp}
\begin{equation}
	\boldsymbol{\mathcal{H}}(\boldsymbol{M}, \boldsymbol{r},t)=\boldsymbol{\mathcal{H}}_{\rm{ex}}(\boldsymbol{M}, \boldsymbol{r},t)+\boldsymbol{\mathcal{H}}_{\rm{an}}(\boldsymbol{M}, \boldsymbol{r},t)+\boldsymbol{\mathcal{H}}_{\rm{dm}}(\boldsymbol{M}, \boldsymbol{r},t).
\end{equation}
It includes the effective field arising from exchange interactions $\boldsymbol{\mathcal{H}}_{\rm{ex}}(\boldsymbol{M}, \boldsymbol{r},t)$, the effective field due to magnetocrystalline anisotropy $\boldsymbol{\mathcal{H}}_{\rm{an}}(\boldsymbol{M}, \boldsymbol{r},t)$, and the demagnetizing field $\boldsymbol{\mathcal{H}}_{\rm{dm}}(\boldsymbol{M}, \boldsymbol{r},t)$ resulting from dipole-dipole interactions.
The dependence of the effective field $\boldsymbol{\mathcal{H}}(\boldsymbol{M}, \boldsymbol{r},t)$ on the magnetization $\boldsymbol{M}$ leads to strong nonlinearities in the LL equation.
Next, we apply certain physical approximations to linearize the LL equation.
Consider applying a sufficiently strong biased magnetic field in the $z$-direction, which will ensure the saturation of magnetization in the magnetic sphere.
Then, the physical field can be expressed as the classical term plus its corresponding fluctuation term
\begin{equation}
	\begin{split}
		\boldsymbol{H}(\boldsymbol{r},t)&=H_0\boldsymbol{e}_z+\boldsymbol{h}(\boldsymbol{r},t), \\
		\boldsymbol{M}(\boldsymbol{r},t)&=M_s\boldsymbol{e}_z+\boldsymbol{m}(\boldsymbol{r},t),
	\end{split}
	\label{FLUA}
\end{equation}
where the fluctuation terms satisfy $\boldsymbol{h}\ll H_0$ and $\boldsymbol{m}\ll M_s$.
By substituting Eq.~(\ref{FLUA}) into the nonlinear Landau-Lifshitz Eq.~(\ref{LLO}), the linearized LL equation is obtained as
\begin{equation}
	\frac{1}{\vert \gamma_e\vert \mu_0 M_s H_0}\partial_t \boldsymbol{m}(\boldsymbol{r},t)-\boldsymbol{e}_z \times \left[\frac{\boldsymbol{m}(\boldsymbol{r},t)}{M_s}-\frac{\boldsymbol{h}(\boldsymbol{r},t)}{H_0}\right]+\left[\boldsymbol{e}_z+\frac{\boldsymbol{m}(\boldsymbol{r},t)}{M_s}\right]\times \frac{\Delta\boldsymbol{H}(\boldsymbol{M}, \boldsymbol{r},t)}{H_0}=0.
	\label{LLA1}
\end{equation}
The higher-order terms of $\boldsymbol{h}/H_0$ and $\boldsymbol{m}/M_s$ have been neglected here, and the first-order term $\boldsymbol{m}/M_s$ satisfies $\boldsymbol{m}\cdot\boldsymbol{e}_z=0$.

Next, we simplify the effective field $\boldsymbol{\mathcal{H}}(\boldsymbol{M}, \boldsymbol{r},t)$.
When the size of the micromagnetic sphere is much larger than the domain wall length, dipole-dipole interactions within the sphere dominate, and the contribution from exchange interactions can be neglected, i.e., $\boldsymbol{\mathcal{H}}_{\rm{ex}}(\boldsymbol{M}, \boldsymbol{r},t)\approx 0$~\cite{2000Aharonip,2009Stancilp}.
As for the magnetocrystalline anisotropy, its contribution to the effective field is a second-order term in $\boldsymbol{m}/M_s$, so it can be neglected~\cite{2000Aharonip,2009Stancilp}.
For a uniformly magnetized ellipsoid, the demagnetizing field arising from dipole-dipole interactions can be written as $\boldsymbol{\mathcal{H}}_{\rm{dm}}=-M_s\boldsymbol{e}_z/3$~\cite{1975Jacksonp,2000Aharonip}.
Based on the above three approximations, the linearized LL equation can be written as~
\begin{equation}
	\begin{aligned}
		-\partial_t m_x\left(\boldsymbol{r},t\right)&=\vert \gamma_e\vert \mu_0\left(H_0-\frac{1}{3}M_s\right)m_y\left(\boldsymbol{r},t\right)-\vert \gamma_e\vert \mu_0M_sh_y\left(\boldsymbol{r},t\right), \\
		\partial_t m_y\left(\boldsymbol{r},t\right)&=\vert \gamma_e\vert \mu_0\left(H_0-\frac{1}{3}M_s\right)m_x\left(\boldsymbol{r},t\right)-\vert \gamma_e\vert \mu_0M_sh_x\left(\boldsymbol{r},t\right).
	\end{aligned}
\end{equation}
Defining the frequencies $\omega_0\equiv\vert\gamma_e \vert \mu_0 H_I$ and $\omega_M\equiv\vert\gamma_e\vert \mu_0 M_s$ with the internal field $H_I\equiv H_0-1/3M_s$, the linearized Landau-Lifshitz equation can be further simplified to~\cite{2009Stancilp}
\begin{equation}
	\begin{aligned}
		\partial_t m_x(\boldsymbol{r},t)&=-\omega_0 m_y(\boldsymbol{r},t)+\omega_M h_y(\boldsymbol{r},t), \\
		\partial_t m_y(\boldsymbol{r},t)&=\omega_0 m_x(\boldsymbol{r},t)-\omega_M h_x(\boldsymbol{r},t).
	\end{aligned}
	\label{LLDD}
\end{equation}

Finally, we consider the magnetostatic approximation $\nabla \times \boldsymbol{h}(\boldsymbol{r},t)\approx 0$.
Under this approximation, (i) the electric field component of the spin wave and $\boldsymbol{h}$ in Maxwell's equations are decoupled and can thus be eliminated.
(ii) The magnetostatic potential is defined as $\boldsymbol{h}(\boldsymbol{r},t)=-\nabla\psi(\boldsymbol{r},t)$.
According to the zero-divergence condition $\nabla\cdot\boldsymbol{b}=0$ and $\boldsymbol{b}(\boldsymbol{r},t)=\mu_0[\boldsymbol{h}(\boldsymbol{r},t)+\boldsymbol{m}(\boldsymbol{r},t)]$, the magnetostatic potential satisfies the equation
\begin{equation}
	\nabla^2 \psi(\boldsymbol{r},t)=\partial_x m_x(\boldsymbol{r},t)+\partial_y m_y(\boldsymbol{r},t).
	\label{PsiIn}
\end{equation}
Outside the magnetic sphere, the scalar field $\psi(\boldsymbol{r},t)$ satisfies the equation
\begin{equation}
	\nabla^2 \psi(\boldsymbol{r},t)=0.
	\label{PsiOu}
\end{equation}

\subsection{The Walker modes}
Here, we focus on the Walker mode in the micromagnetic sphere.
Similar to the quantization of the electromagnetic field, by using the eigenmodes $\boldsymbol{m}_\beta(\boldsymbol{r})$ of the magnetization and the eigenmodes $\boldsymbol{h}_\beta(\boldsymbol{r})=-\nabla \psi_\beta(\boldsymbol{r})$ of the magnetic field, the corresponding physical field can be represented as~\cite{1959Fletcherp687,1977Roeschmannp11}
\begin{equation}
	\begin{split}
		\boldsymbol{m}\left(\boldsymbol{r},t\right)&=\sum_{\beta}\left[s_{\beta} \boldsymbol{m}_{\beta}\left(\boldsymbol{r}\right)e^{-i\omega_{\beta}t}+c.c\right], \\
		\boldsymbol{h}\left(\boldsymbol{r},t\right)&=\sum_{\beta}\left[s_{\beta} \boldsymbol{h}_{\beta}\left(\boldsymbol{r}\right)e^{-i\omega_{\beta}t}+c.c\right],
	\end{split}
	\label{MHEigen}
\end{equation}
where $s_\beta$, $\omega_\beta$ and $\beta$ denote complex amplitude, eigenfrequency, and mode indices, respectively.
Then the linearized LL equations can be obtained that
\begin{equation}
	\begin{aligned}
		i\omega m_x (\boldsymbol{r})&=\omega_M\partial_y\psi(\boldsymbol{r})+\omega_0 m_y(\boldsymbol{r}), \\
		i\omega m_y (\boldsymbol{r})&=-\omega_M\partial_x\psi(\boldsymbol{r})-\omega_0 m_x(\boldsymbol{r}).
	\end{aligned}
	\label{LLDD2}
\end{equation}
By inserting equation (\ref{LLDD2}) into the equation for the magnetostatic potential, we obtain the equation
\begin{subequations}
	\begin{equation}
		\left(1+\chi_p\right)\left(\partial_x^2+\partial_y^2\right)\psi_{\rm{in}}(\boldsymbol{r})+\partial_z^2 \psi_{\rm{in}}(\boldsymbol{r})=0,
		\label{SWEQIN}
	\end{equation}
	\begin{equation}
		\nabla^2 \psi_{\rm{out}}(\boldsymbol{r})=0,
		\label{SWEQOUT}
	\end{equation}
\end{subequations}
satisfied by the potential $\psi_\mathrm{in}(\boldsymbol{r},t)$ inside the magnetic sphere and the potential $\psi_\mathrm{out}(\boldsymbol{r},t)$ outside the sphere.
Here, the diagonal element of the Polder susceptibility tensor is defined as $\chi_p(\omega)=\omega_M\omega_0/(\omega_0^2-\omega^2)$~\cite{2009Stancilp}.

Next, we solve the equations for the magnetostatic potential satisfied inside and outside the sphere, respectively.
Since we are considering spin wave modes in a micromagnetic sphere, it is more convenient to use spherical coordinates.
First, the magnetostatic potential outside the sphere satisfies Eq.~(\ref{SWEQOUT}).
The general solution for the magnetostatic potential outside the magnetic sphere in spherical coordinates is
\begin{equation}
	\psi_{\rm{out}}(\boldsymbol{r})=\sum_{l,m}\left[\frac{A_{l,m}}{r^{l+1}}+B_{l,m}r^l\right]Y_l^m(\theta,\phi),
\end{equation}
where $Y_l^m(\theta,\phi)$ is a spherical harmonic function in spherical coordinates.
The $A_{l,m}$ and $B_{l,m}$represent the expansion coefficients, which can be determined from the boundary conditions.
For the interior of the magnetic sphere, the magnetostatic potential satisfies Eq.~(\ref{SWEQIN}).
For convenience, we introduce a new set of coordinates $\{\xi,\eta,\phi\}$, defined by $x=\sqrt{\chi_p}R\sqrt{\xi^2-1}\sin\eta \cos \phi$, $y=\sqrt{\chi_p}R\sqrt{\xi^2-1}\sin\eta \sin \phi$, and $z=\sqrt{\chi_p/(1+\chi_p)}R\xi\cos\eta$.
At the surface of the sphere, the coordinates $\{\xi,\eta,\phi\}$  simplify to $\xi\rightarrow\xi_0=\sqrt{(1+\chi_p)/\chi_p}$, $\eta\rightarrow \theta$, and $\phi\rightarrow\phi$.
In orthogonal coordinates $\{\xi,\eta,\phi\}$, the generalized solution for the magnetostatic potential inside the sphere can be expressed in terms of the associated Legendre polynomials and spherical harmonic functions~\cite{1959Fletcherp687,1977Roeschmannp11}
\begin{equation}
	\psi_{\rm{in}}(\boldsymbol{r})=\sum_{l,m}C_{l,m}P_l^m(\xi)Y_l^m(\eta,\phi),
\end{equation}
where $C_{l,m}$ is the expansion coefficient.

A detailed analysis of the use of boundary conditions to determine the expansion coefficients follows.
(i) the magnetostatic potential is convergent at infinity ($\psi_{\rm{out}}(\boldsymbol{r}\rightarrow \infty)=0$), then we have $B_{l,m}=0$.
(ii) By applying the continuity of the tangential component of the field $\boldsymbol{h}$ on the surface, it follows that $\psi_{\rm{out}}\vert _{r=R}=\psi_{\rm{in}}\vert _{\xi=\xi_0}$.
Consequently, we obtain $A_{l,m}=C_{l,m}P_l^m(\xi_0)R^{l+1}$.
(iii) By applying the continuity condition for the normal component of the field $\boldsymbol{b}$~\cite{1959Fletcherp687}, one can obtain
\begin{equation}
	\partial_r\psi_{\rm{out}}\vert_{r=R}=\frac{\xi_0}{R}\partial_{\xi}\psi_{\rm{in}}\vert_{r=R}-i\frac{\kappa_p}{R}\partial_{\phi}\psi_{\rm{in}}\vert_{r=R},
\end{equation}
where the off-diagonal element of the Polder susceptibility tensor is given by $i\kappa_p(\omega)$, with $\kappa_p(\omega)=\omega_M\omega/(\omega_0^2-\omega^2)$.
Combining the above boundary conditions with $A_{l,m}=C_{l,m}P_l^m(\xi_0)R^{l+1}$, the equation satisfied by the eigenfrequency of Walker's mode is derived as~\cite{1959Fletcherp687,1977Roeschmannp11}:
\begin{equation}
	\xi_0(\omega)\frac{{P_l^{\prime}}^m\left[\xi_0(\omega)\right]}{P_l^m\left[\xi_0(\omega)\right]}+m\kappa_p(\omega)+l+1=0.
\end{equation}
From this equation, it can be concluded that the eigenfrequencies are independent of the size of the micromagnetic sphere, as the equation does not depend on the sphere's size.
Spin-wave modes are characterized by three indices $\{n,l,m\}$.
It is worth noting that there are nonphysical solutions to this equation, such as $l=0$.
Here, we focus on a special spin-wave mode in the micromagnetic sphere, the Kittel mode, in which all magnetic moments precess with the same amplitude and phase.

\subsection{Quantization of the spin wave modes}
This section will provide a detailed discussion of the quantization of spin-wave modes.
The micromagnetic energy functional  of the Walker mode can be expressed as
\begin{equation}
	E_m\left(\{\boldsymbol{m}\},\{\boldsymbol{h}\}\right)=\frac{\mu_0}{2}\int dV \boldsymbol{m}(\boldsymbol{r},t)\cdot\left[\frac{H_I}{M_s}\boldsymbol{m}(\boldsymbol{r},t)-\boldsymbol{h}(\boldsymbol{r},t)\right].
	\label{EMO}
\end{equation}
and it can be shown that the LL equations are reproduced using this energy functional (\ref{EMO}).
By utilizing the linear LL equation (\ref{LLDD}), we can obtain
\begin{subequations}
	\begin{equation}
		h_x(\boldsymbol{r},t)=-\frac{1}{\omega_M}\partial_t m_y(\boldsymbol{r},t)+\frac{\omega_0}{\omega_M}m_x(\boldsymbol{r},t),
		\label{hx}
	\end{equation}
	\begin{equation}
		h_y(\boldsymbol{r},t)=\frac{1}{\omega_M}\partial_t m_x(\boldsymbol{r},t)+\frac{\omega_0}{\omega_M}m_y(\boldsymbol{r},t).
		\label{hy}
	\end{equation}
\end{subequations}
By submitting Eqs.~(\ref{hx}) and ~(\ref{hy}) into the energy functional (\ref{EMO}), we can get
\begin{equation}
	E_m\left(\{\boldsymbol{m}\}\right)=\frac{1}{2\vert \gamma_e \vert M_s}\int dV \{m_x(\boldsymbol{r},t)\partial_t m_y (\boldsymbol{r},t)-m_y(\boldsymbol{r},t) \partial_t m_x(\boldsymbol{r},t)\}.
	\label{EM1}
\end{equation}
Expanding the energy functional (\ref{EMO}) in terms of the eigenmode (\ref{MHEigen}) yields
\begin{equation}
	E_m\left(\{\boldsymbol{m}\}\right)=\frac{\Lambda_{\beta}}{2\vert \gamma_e \vert \hbar M_s} \sum_{\beta} \hbar\omega_{\beta}\left(s_{\beta}s_{\beta}^* + s_{\beta}^*s_{\beta}\right),
	\label{EM2}
\end{equation}
where $\Lambda_{\beta}=\int dV 2 {\rm{Im}} [m_{\beta,y}(\boldsymbol{r})m_{\beta,x}^*(\boldsymbol{r})]$~\cite{1957Walkerp390,2006Millsp16}.
It can be seen that the energy functional of the Walker mode is the sum of the energies of a series of harmonic oscillators.
Similar to the harmonic oscillator quantization, we quantize the amplitudes into magnon operators $s_{\beta}\rightarrow\hat{s}_{\beta}$ and $s_{\beta}^*\rightarrow\hat{s}_{\beta}^{\dagger}$.
A suitable eigenfunction
\begin{equation}
	\left[
	\begin{array}{l}
		\boldsymbol{m}_{\beta}(\boldsymbol{r}) \\
		\boldsymbol{h}_{\beta}(\boldsymbol{r})
	\end{array}
	\right] \rightarrow M_{\rm{zpf}}
	\left[
	\begin{array}{l}
		\widetilde{\boldsymbol{m}}_{\beta}(\boldsymbol{r}) \\
		\widetilde{\boldsymbol{h}}_{\beta}(\boldsymbol{r})
	\end{array}
	\right]
\end{equation}
is chosen to eliminate the constant factor $\Lambda_{\beta}/(\vert \gamma_e\vert\hbar M_s)$ in the energy functional (\ref{EM2}), where $M_K$ represents the zero-point magnetization.

Using the commutation relation $[\hat{s}_{\beta},\hat{s}_{\beta}^{\dagger}]=1$ for the bosonic operator, the quantized field operator and Hamiltonian are given by
\begin{subequations}
	\begin{equation}
		\hat{\boldsymbol{m}}(\boldsymbol{r})=\sum_{\beta} M_{\rm{zpf}} \left(\widetilde{\boldsymbol{m}}_{\beta}(\boldsymbol{r})\hat{s}_{\beta}+h.c.\right),
	\end{equation}
	\begin{equation}
		\hat{\boldsymbol{h}}(\boldsymbol{r})=\sum_{\beta} M_{\rm{zpf}} \left(\widetilde{\boldsymbol{h}}_{\beta}(\boldsymbol{r})\hat{s}_{\beta}+h.c.\right),
	\end{equation}
	\begin{equation}
		\hat{H}=\sum_{\beta} \hbar\omega_{\beta}\left(\hat{s}_{\beta}^{\dagger}\hat{s}_{\beta} + \frac{1}{2}\right).
	\end{equation}
\end{subequations}

\subsection{The Kittel mode}
Here, we focus on the Kittel mode in the YIG magnetic sphere, whose mode function is given by
\begin{equation}
	\widetilde{\boldsymbol{m}}_{K}=\boldsymbol{e}_x+i\boldsymbol{e}_y.
\end{equation}
The Kittel mode corresponds to the constant factor $\Lambda_{\beta}=2V_K$ and the zero-point magnetization $M_K\equiv M_{\rm{zpf}}=\sqrt{\gamma_e \hbar M_s /(2V_K)}$.
The symbol $M_s$ represents the saturation magnetization of the YIG sphere.
The quantized magnetization operator can then be written as
\begin{equation}
	\hat{\boldsymbol{m}}=M_K\left(\widetilde{\boldsymbol{m}}_K\hat{s}_K+\widetilde{\boldsymbol{m}}_K^*\hat{s}_K^\dagger\right).
\end{equation}
The free Hamiltonian of the Kittel mode is given by
\begin{equation}
	\hat{H}_K=\omega_K \hat{s}_K^{\dagger}\hat{s}_K,
\end{equation}
where the resonant frequency is $\omega_K=\gamma_e B_K$.
The symbol $B_K$ represents the bias magnetic field.

\section{\label{eNe}The trapped electron}
\begin{figure}
	\centering
	\includegraphics[width=0.2\textwidth]{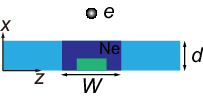}
	\caption{\textbf{Schematic diagram of an eNe-trapped electronic device.}}
	\label{SMFig1}
\end{figure}

In this section, we will provide a detailed analysis of the energy level structure of confined electrons [\ref{SMFig1}].
In the trapped electron platform, electrons are confined above the QLS, which exhibits three distinct degrees of freedom: the Rydberg state, orbital state, and spin.
In the subsequent analysis, we concentrate on the eNe Platform
In this platform, the electrons are confined above the solid-state Ne, which enables longer coherence times compared to the eHe confined electron platform.

\subsection{\label{TRS}The Rydberg state}
In the gravitational direction, the electron can be stably confined in the $z$-direction due to the attraction of the electron's image charge in the QLS and the repulsion of the Pauli potential barrier.
The electron is subjected to a potential of
\begin{equation}
	V =
	\begin{cases}
		-\frac{Ze^2}{x}, & x > 0 \\
		\infty, & x \leq 0
	\end{cases}
\end{equation}
where the effective charge $Z=(\epsilon - 1)/[4(\epsilon + 1)]$ and the dielectric constant $\epsilon$.
The energy levels of the confined electron in the $z$ direction resemble those of a hydrogen atom.
Consequently, the energy spectrum can be expressed as
\begin{equation}
	E_n=-R_{\infty}\left(\frac{\Lambda}{4}\right)^2\frac{1}{n^2},~R_\infty=\frac{m_e e^4}{8\varepsilon_0^2 h^2}
\end{equation}
with the corresponding wave function given by
\begin{equation}
	\psi_n(\rho)=C\rho\exp\left(-\frac{\rho}{2}\right)\mathcal{L}_{n-1}^{(1)}\left(\rho\right),
\end{equation}
where $\mathcal{L}_{n-1}^{(1)}\left(\rho\right)$ is the generalized Laguerre polynomial, $\rho\equiv\alpha x$, and $\alpha\equiv\sqrt{8m_e\vert E_n\vert/\hbar^2}$.
Here, $C$ is the normalization constant of the wave function, and $m_e$ represents the mass of the electron.
Theoretical calculations show that the energies of the ground and first excited states are $E_1 = -2431~\mathrm{GHz}$ and $E_2 =-607.8~\mathrm{GHz}$, respectively, with an energy separation of $\Delta E = 1837.27~\mathrm{GHz}$.
However, the resonance frequency of the system studied here is on the order of several gigahertz, significantly lower than the excitation frequency associated with motion along the $x$-axis.
In addition, the energy spacing between the ground and first excited states of the electron’s motion in the $x$ direction is on the order of THz, far exceeding the thermal energy at millikelvin temperatures, so this degree of freedom is effectively frozen and its dynamics can be neglected~\cite{2022ZhouP4650,2024ZhouP116122}.
Therefore, within the frequency range of interest, the electron’s motion along the $x$-axis can be reasonably assumed to be frozen in the ground state and thus neglected.

\subsection{\label{TOS}The orbital state}
The in-plane motion of the electrons can be confined by the potential created by the electrodes.
The current experimental focus is on generating a confinement potential with a high aspect ratio, so that the motion of the electron in one of the directions can be cooled in the ground state, thus allowing the consideration of motion in only one of the directions.
As illustrated in Fig.~\ref{SMFig1}, we focus solely on the electron's motion in the $z$ direction.
We assume that the electron is confined in a harmonic potential along the $z$-direction, with the corresponding Hamiltonian expressed as~\cite{2010SchusterP4050340503}
\begin{equation}
	\hat{H}_z=\frac{\hat{p}_z^2}{2m_e}+\frac{1}{2}m_e\omega_z^2 z^2,
\end{equation}
where the resonance frequency is $\omega_z/2\pi=\sqrt{eV_0/(m_e W^2)}$ and the applied voltage is $V_0=V_c \exp (-2\pi d/W)$.
Introducing the creation and annihilation operator
\begin{equation}
	\hat{z}=\sqrt{\frac{\hbar}{2m_e\omega_a}}\left(\hat{a}+\hat{a}^\dagger\right)=z_0\left(\hat{a}+\hat{a}^\dagger\right),~\hat{p}_z=\sqrt{\frac{\hbar m_e \omega_a}{2}}\left(\hat{a}-\hat{a}^\dagger\right),
\end{equation}
the quantized Hamiltonian is $\hat{H}_z=\hbar \omega_z \hat{a}^\dagger \hat{a}$.

\subsection{\label{DMS}The decoherence of the motional and spin states}
This section analyzes the coherence of the electron's motion and spin states.

\subsubsection{The motional states}
Both liquid helium (He) and solid neon (Ne) serve as substrate materials for constructing trapped electron platforms.
We first examine the dissipation mechanism of the electron's motion state when liquid He serves as the substrate.
Specifically, the dissipation of the electron's motion state in a liquid He-based platform occurs through two mechanisms: electrical fluctuations on the electrodes and excitations of the liquid He~\cite{2010SchusterP4050340503}.

The electrode-induced dissipation mechanism consists of two types.
The first is dissipation due to the spontaneous radiation of electrons into free space, which can be effectively suppressed by precisely tuning the microwave environment.
The second is the decoherence effect caused by the voltage fluctuation at the electrodes.
This decoherence effect is attributed to the change in the trapped potential due to voltage drift, which in turn causes a shift in the trapped frequency.
The generation of voltage fluctuation is mainly affected by several factors: the drift of the voltage source, thermal Josephson voltage noise, and local $1/f$ charge noise.
Among them, the voltage source drift can be eliminated by compensating methods; the effect of thermal Josephson voltage noise is negligible at the very low temperature of $10~\mathrm{mK}$; and the local $1/f$ charge noise can be effectively suppressed by applying a large capacitance to ground.

The dissipation mechanism of liquid He excitations arises from ripplon excitations, including linear single-ripplon and second-order two-ripplon excitations.
The contribution of single-ripplon excitation to dissipation is minimal due to the large separation between the electrons and the liquid He surface.
For double-ripplon excitation, the mechanism involves the simultaneous interaction of two ripplons with nearly opposite momenta and electrons, with a corresponding decay rate estimated at $1~\mathrm{kHz}$~\cite{2010SchusterP4050340503}.
The main dissipation channel for liquid He excitations is phonon decay.
The mechanism is analogous to piezoelectric coupling in semiconductors.
The decay rate of this process is approximately $30~\mathrm{kHz}$~\cite{2010SchusterP4050340503}.
Simultaneously, the coupling of electrons to ripplons induces a decoherence effect, which is significantly reduced at low temperatures.
The decoherence rate is estimated to be $2~\mathrm{kHz}$ at $50~\mathrm{mK}$~\cite{2010SchusterP4050340503} and can be further reduced at lower temperatures.

In summary, based on the theoretical analysis of the dissipation mechanisms, the coherence time of the electron's motion state trapped on liquid He is expected to exceed $15~\mathrm{\mu s}$, corresponding to a decoherence rate below $0.067~\mathrm{MHz}$~\cite{2010SchusterP4050340503}.
However, the experimentally measured decoherence rate is as high as $77~\mathrm{MHz}$, approximately $1000$ times higher than the theoretical value~\cite{2019KoolstraP53235323}.
This significant discrepancy is likely primarily due to additional dissipation caused by the fluctuations of liquid He induced by the pulse tube refrigerator.

Replacing the substrate material from liquid He to solid Ne eliminates liquid He excitations---the main dissipation mechanism for electron motion state decoherence---significantly improving the electron's coherence time.
Theoretical predictions suggest that the coherence time of electrons trapped on the surface of solid Ne can reach milliseconds~\cite{2005ZavyalovP415420}, and experimental results confirm this prediction~\cite{2022ZhouP4650,2024ZhouP116122}.
Without any optimization, the relaxation time of the electron's motion state reaches $15~\mathrm{\mu s}$, while the phase coherence time is $200~\mathrm{ns}$~\cite{2022ZhouP4650}.
In solid-state Ne substrates, the dissipation mechanism of the electron's motion state arises from background charge noise due to residual electrons and the imperfect surface properties of solid-state Ne. For example, surface rough and a porous structure may lead to highly movable neon atoms, which generate a time-dependent trapped potential for the electrons.
Experimentally, the coherence of the electron motion state is further optimized by three important improvements: annealing the solid-state Ne to form a higher quality surface; stabilizing the electron confinement potential so as to minimize the background charge noise; and tuning the motion state to an operating state that is insensitive to the charge noise.
As a result of these improvements, both the relaxation time and coherence time of the motion state of the solid-state Ne surface-trapped electrons are improved to $0.1~\mathrm{ms}$, which is significantly better than the experimental results before the optimization~\cite{2024ZhouP116122}.
This progress provides an important experimental basis for a solid-state Ne-based quantum information processing platform.

\subsubsection{The spin states}

\begin{figure}
	\centering
	\includegraphics[width=0.75\textwidth]{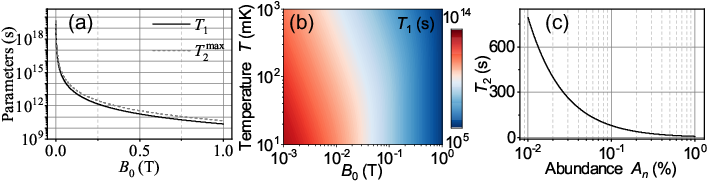}
	\caption{\textbf{Dissipation mechanisms in the spin state of a trapped electron.} (a) Spontaneous radiation of electrons into the vacuum, (b) Phonon-induced fluctuations in the Ne diamagnetic susceptibility, (c) Spin-spin interactions between electron and nuclear spins.}
	\label{SMFig2}
\end{figure}

Electron spin states exhibit superior coherence properties compared to electron motion states.
For electron spin states trapped on the surface of liquid He, the dissipation mechanisms include spin-orbit coupling, fluctuating magnetic fields, and magnetic flux noise.
Theoretical analysis indicates that even under the most unfavorable conditions, the decoherence rate of the electron spin state remains as low as $200~\mathrm{Hz}$~\cite{2010SchusterP4050340503}, which further highlights its potential advantages in quantum information processing.

On solid-state Ne surfaces, the coherence of electron spin states has been analyzed in detail~\cite{2022ChenP4501645016}.
The dissipation mechanisms include spontaneous radiation of electrons into the vacuum, phonon-induced fluctuations in  the Ne diamagnetic susceptibility, thermal currents and local spin-density fluctuations in metallic electrodes, and spin-spin interactions between electron and nuclear spins.
Spontaneous radiation of electron spin into the vacuum occurs through magnetic dipole radiation.
If the electron is initialized in the excited state, its relaxation rate to the ground state is given by
\begin{equation}
	\frac{1}{T_1}=\frac{2\mu_0 g_s^2 \mu_B^2 \omega_s^3}{12\pi \hbar c^3},
\end{equation}
where $\mu_0$ is the vacuum permeability, $g_s$ is the electron g-factor, $\mu_B$ is the Bohr magneton, $\omega_s$ is the electron spin-transition frequency, and $c$ is the speed of light.
According to the relation $1/T_2=1/(2T_1)+1/T_\varphi$, the upper bound on the coherence time satisfies $T_2\leq 2T_1\equiv T_2^\mathrm{max}$.
Here, $T_\varphi$ is the pure dephasing time.
It is evident that the dissipation mechanism due to magnetic dipole radiation depends on the transition frequency of the electron spin state.
Figure~\ref{SMFig2}(a) illustrates the relationship between the relaxation time $T_1$ and the applied magnetic field $B_0$, indicating that the contribution of magnetic dipole radiation to the dissipation of the electron spin state is negligible.

The phonon-induced dissipation mechanism arises from variations in the diamagnetic susceptibility of solid-state Ne.
When a uniform magnetic field is applied to electrons and solid-state Ne, the diamagnetism of Ne atoms induces magnetization, creating a magnetization current at the surface.
This current generates a magnetic field that affects the electron spin state.
Additionally, thermal fluctuations of bulk phonons in solid-state Ne modify the volume magnetic susceptibility by altering Ne's mass density, which in turn causes fluctuations in the magnetization current.
These fluctuations lead to variations in the magnetic field acting on the electron spins, resulting in relaxation and decoherence of the electron spin state.
The relaxation rate of electrons due to this process is given by
\begin{equation}
	1/T_1=\frac{\chi^2 \mu_B^2 B_0^2 \omega_s}{64\pi\hbar\rho v_s^3 d^2} \frac{1}{1-\exp(-\frac{\hbar \omega_s}{k_B T})},
\end{equation}
where $\chi=-6.74\times 10^5$ is the volume magnetic susceptibility of solid Ne and $\rho=m_\mathrm{Ne}n$ represents the Ne mass density.
Here, $m_\mathrm{Ne}=3.35\times 10^{-28}~\mathrm{kg}$ is the mass of a Ne atom, $n$ is the particle number density, and $v_s=704~\mathrm{m/s}$ denotes the speed of sound in solid Ne.
Notably, this process results in a zero electron decoherence rate, implying that the coherence time satisfies $T_2=2T_1$~\cite{2022ChenP4501645016}.
Figure~\ref{SMFig2}(b) illustrates the dependence of the relaxation time $T_1$​ on the spin transition frequency and temperature.
As shown in the Fig.~\ref{SMFig2}(b), the contribution of phonon-induced dissipation to the relaxation of the electron spin state is minimal and can be considered negligible.
This result confirms that electron spin states on the solid-state Ne surface retain excellent coherence properties even under the influence of phonon-induced dissipation, which can be disregarded.

When superconducting Al or Nb films are used for electrode, the dissipation mechanism in the electrodes due to thermal currents and localized spin-density fluctuations is effectively eliminated.
In contrast, if the electrodes are made of conventional metals, fluctuations in thermal currents induce variations in the magnetic field, leading to the decoherence of electron spin states.
Additionally, electron spin density fluctuations in metals generate magnetic field fluctuations, leading to decoherence in the electron spin state.
However, when the electrodes are composed of superconducting thin films (e.g., Al or Nb), this decoherence mechanism is exponentially suppressed.
This is because, in s-wave superconductors like Al and Nb, the carriers are Cooper pairs with zero total spin.
The frequency of quasiparticle excitations (tens of $\mathrm{GHz}$) is significantly higher than both the temperatures ($10~\mathrm{mK}$) and the operating frequencies (a few $\mathrm{GHz}$) typically used in quantum computing experiments.
As a result, dissipation mechanisms induced by thermal currents and spin-density fluctuations can be entirely neglected in superconducting electrodes.

Thus, for the electron spin state, the primary dissipation mechanism originates from the spin-spin interaction between the electron spin and the Ne nucleus.
Ne in nature is composed of three isotopes: $^{20}\mathrm{Ne}$ ($90.48\%$), $^{21}\mathrm{Ne}$ ($0.27\%$), and $^{22}\mathrm{Ne}$ ($9.25\%$).
Among these, $^{20}\mathrm{Ne}$ ($90.48\%$) and $^{22}\mathrm{Ne}$ ($9.25\%$) have a nuclear spin of zero and, therefore, do not affect the coherence of electron spin states.
However, $^{21}\mathrm{Ne}$ ($0.27\%$), with a natural abundance of $0.27\%$, has a nuclear spin of $3/2$, and its interaction with the electron spin induces dissipation.
For solid Ne, the dissipation due to nuclear spin can be described by the phenomenological expression:
\begin{equation}
	T_2=\left(A_n=0.27\%\right)=c^\prime \vert g_n\vert^{-1.6} I_n^{-1.1} \left(A_n n\right)^{-1},
\end{equation}
where $c^\prime=1.5\times 10^{24}~\mathrm{m^{-3}~s}$ is an isotope-independent constant, $g_n=-0.44$ is the nuclear spin g-factor of $^{21}\mathrm{Ne}$, and $I_n=3/2$ is the nuclear spin quantum number of $^{21}\mathrm{Ne}$.
$A_n$ represents the abundance of $^{21}\mathrm{Ne}$.
Figure~\ref{SMFig2}(c) illustrates the dependence of the electron spin coherence time $T_2$ on the abundance $A_n$ of $^{21}\mathrm{Ne}$.
As shown in the Fig.~\ref{SMFig2}(c), a lower $A_n$ corresponds to a longer coherence time.
With commercially purified Ne and the application of Hahn echoes, the coherence time reaches up to $81~\mathrm{s}$, corresponding to a decoherence rate of $\gamma_s=0.01~\mathrm{Hz}$~\cite{2022ChenP4501645016}.

\section{\label{CM}The coupling mechanism}
This section provides a detailed discussion of the interaction mechanism between magnons in the YIG micromagnetic sphere and the eNe quantum platform.
The interaction between the magnon and the eNe quantum platform is mediated by the stray field surrounding the YIG micromagnetic sphere.
The interaction form between the two and the corresponding coupling strength will be calculated in detail in the following section.

\subsection{\label{TQMF}The quantized magnetic fields}
For a YIG magnetic sphere with a magnetic moment $\boldsymbol{\mu}$, the magnetic field at position $\boldsymbol{r}$ is described by the magnetic dipole model
\begin{equation}
	\boldsymbol{B}=\frac{\mu_0}{4\pi}\left[\frac{3\boldsymbol{r}\left(\boldsymbol{\mu}\cdot\boldsymbol{r}\right)}{r^5}-\frac{\boldsymbol{\mu}}{r^3}\right],
	\label{DM}
\end{equation}
where $\boldsymbol{\mu}=\boldsymbol{m}V_K$ represents the magnetic moment of the YIG micromagnetic sphere, and $V_K=4\pi R_K^3/3$ denotes its volume.
The symbol $\mu_0$ is the vacuum permeability.
Here, we define a coordinate system with the equilibrium position of the electron moving along the $z$-axis as the origin.
The position coordinates of the magnetic sphere and the electron are $(H,0,0)$ and $(0,0,z)$, respectively.
Consequently, the position vector from the field point to the source point is $\boldsymbol{r}=(-H,0,z)$.
Here, $H=R_K+d_K$, where $R_K$ is the radius of the YIG micromagnetic sphere and $d_K$ is the distance from the surface of the YIG micromagnetic sphere to the electron.
Next, we introduce the quantized magnetization operator $\hat{\boldsymbol{m}}=M_K(\widetilde{\boldsymbol{m}}_K\hat{s}_K+\widetilde{\boldsymbol{m}}_K^*\hat{s}_K^\dagger)$, where $\widetilde{\boldsymbol{m}}_K=\boldsymbol{e}_x + i\boldsymbol{e}_y$ represents the mode function corresponding to the Kittel mode.
Substituting the quantized magnetization operator into Eq.~(\ref{DM}) yields the quantized magnetic field $\hat{\boldsymbol{B}}=(\hat{B}_x,\hat{B}_y,\hat{B}_z)$, where
\begin{equation}
	\begin{aligned}
		\hat{B}_x&=\frac{\mu_0 V_K M_K}{4\pi}\frac{2H^2-z^2}{r^5}\left(\hat{s}_K+\hat{s}_K^\dagger\right),\\
		\hat{B}_y&=\frac{\mu_0 V_K M_K}{4\pi}\left[-\frac{i\left(\hat{s}_K-\hat{s}_K^\dagger\right)}{r^3}\right], \\
		\hat{B}_z&=\frac{\mu_0 V_K M_K}{4\pi}\left[-\frac{3H\left(\hat{s}_K+\hat{s}_K^\dagger\right)}{r^5}z\right].
	\end{aligned}
	\label{ME}
\end{equation}
The symbol $M_K$ represents the zero-point magnetization, and $r=\sqrt{z^2+H^2}$ denotes the distance between the field point and the source point.
Near the equilibrium position $z=0$ (i.e., for $\vert z/H\vert \ll 1$), the magnetic field is expanded via a second-order Taylor series, yielding
\begin{equation}
	\hat{B}_x=B_c\left(2-\frac{6z^2}{H^2}\right)\hat{m}_x,~\hat{B}_y=B_c\left(-1+\frac{3z^2}{2H^2}\right)\hat{m}_y,~\hat{B_z}=-\frac{3B_c z}{H}\hat{m}_x,
	\label{MEAPP}
\end{equation}
where $B_c=\mu_0 V_K M_K/(4\pi H^3)$. Here, $\hat{m}_{x,y,z}$ denote the three components of the quantized magnetization operator $\hat{\boldsymbol{m}}$.

\subsection{\label{TIH}The interaction Hamiltonian}
The interaction Hamiltonian between the quantized magnetic field and the electron spin can be expressed as
\begin{equation}
	\hat{H}_\mathrm{KE}=-\gamma_e\hat{\boldsymbol{B}}\cdot\hat{\boldsymbol{S}}=-\frac{\gamma_e}{2}\left(\hat{B}_x\hat{\sigma}_x+\hat{B}_y\hat{\sigma}_y+\hat{B}_z\hat{\sigma}_z\right),
	\label{HIntOrigin}
\end{equation}
where $\gamma_e$ is the gyromagnetic ratio, $\hat{\boldsymbol{S}}=1/2\hat{\boldsymbol{\sigma}}$ is the electron spin operator, and $\hat{\boldsymbol{\sigma}}=(\hat{\sigma}_x,\hat{\sigma}_y,\hat{\sigma}_z)$ represents the Pauli matrices.
Substituting the individual magnetic field components into Eq.~(\ref{HIntOrigin}), the interaction Hamiltonian can be expressed as
\begin{equation}
	\hat{H}_\mathrm{KE}=-\frac{\gamma_e B_c}{2}\left(2-\frac{6\hat{z}^2}{H^2}\right)\hat{m}_x\hat{\sigma}_x - \frac{\gamma_e B_c}{2}\left(-1+\frac{3\hat{z}^2}{2H^2}\right)\hat{m}_y\hat{\sigma}_y+\frac{\gamma_e B_c}{2}\frac{3B_c\hat{z}}{H}\hat{m}_x\hat{\sigma}_z.
\end{equation}
By introducing the phonon creation and annihilation operators $\hat{a}^\dagger$ and $\hat{a}$, the position coordinate can be expressed as $\hat{z}=z_0(\hat{a}+\hat{a}^\dagger)$.
Based on this, the interaction Hamiltonian can be quantized as
\begin{equation}
	\begin{split}
		\hat{H}_\mathrm{KE}=&-g_\mathrm{Tx}(\hat{s}_K+\hat{s}_K^\dagger)\hat{\sigma}_x+g_\mathrm{Ty}(\hat{s}_K-\hat{s}_K^\dagger)(\hat{\sigma}_+ -\hat{\sigma}_-)+g_L(\hat{a}+\hat{a}^\dagger)(\hat{s}_K+\hat{s}_K^\dagger)\hat{\sigma}_z \\ &+g_x^{(2)}(\hat{a}+\hat{a}^\dagger)^2(\hat{s}_K+\hat{s}_K^\dagger)\hat{\sigma}_x-g_y^{(2)}(\hat{a}+\hat{a}^\dagger)^2(\hat{s}_K-\hat{s}_K^\dagger)(\hat{\sigma}_+-\hat{\sigma}_-)
	\end{split}
\end{equation}
with the coupling strengths defined as $g_\mathrm{Tx}=2\mathcal{C}$, $g_\mathrm{Ty}=\mathcal{C}$, $g_L=3\mathcal{C}\epsilon_0$, $g_x^{(2)}=6\mathcal{C}\epsilon_0^2$, and $g_y^{(2)}=3\mathcal{C}\epsilon_0^2/2$.
Additionally, the constant $\mathcal{C}$ is defined as $\mathcal{C}=\gamma_e B_c/2$, and the parameter $\epsilon_0=z_0/H$ satisfies the condition $\epsilon_0 \ll 1$.

Next, we describe the derivation of the nonlinear tripartite interaction.
We assume that the magnons and spin modes are in a non-resonant regime, satisfying the detuning condition $\vert \omega_K-\omega_s\vert \gg g_\mathrm{Tx},g_\mathrm{Ty}$.
First, we apply the rotating-wave approximation (RWA), neglecting the first two terms of the Hamiltonian $\hat{H}_\mathrm{KE}$ (i.e., the high-frequency oscillation terms), to obtain the simplified effective Hamiltonian
\begin{equation}
	\hat{H}_\mathrm{KE}=g_L(\hat{a}+\hat{a}^\dagger)(\hat{s}_K+\hat{s}_K^\dagger)\hat{\sigma}_z +g_x^{(2)}(\hat{a}+\hat{a}^\dagger)^2(\hat{s}_K+\hat{s}_K^\dagger)\hat{\sigma}_x-g_y^{(2)}(\hat{a}+\hat{a}^\dagger)^2(\hat{s}_K-\hat{s}_K^\dagger)(\hat{\sigma}_+-\hat{\sigma}_-).
\end{equation}
Subsequently, the system frequency is further adjusted to induce detuning between the motional state and the magnon mode.
By applying the RWA again, the Hamiltonian is simplified to
\begin{equation}
	\hat{H}_\mathrm{KE}=g_x^{(2)}(\hat{a}+\hat{a}^\dagger)^2(\hat{s}_K+\hat{s}_K^\dagger)\hat{\sigma}_x-g_y^{(2)}(\hat{a}+\hat{a}^\dagger)^2(\hat{s}_K-\hat{s}_K^\dagger)(\hat{\sigma}_+-\hat{\sigma}_-).
	\label{HKEThreeBody}
\end{equation}
At this stage, the system Hamiltonian retains only higher-order nonlinear interaction terms.
We then analyze the resonance conditions of these nonlinear terms.
By expanding the interaction Hamiltonian Eq.~(\ref{HKEThreeBody}), the results summarized in Table~\ref{HKEALL} are obtained.
After neglecting the fast-oscillating terms No.~2 and No.~7, three distinct resonance conditions, $\Delta_1=2\omega_z+\omega_K-\omega_s$, $\Delta_2=2\omega_z-\omega_K-\omega_s$, and $\Delta_3=2\omega_z-\omega_K+\omega_s$, are identified, giving rise to three types of nonlinear tripartite couplings with the corresponding effective Hamiltonians
\begin{subequations}
	\begin{align}
		\Delta_1&=0,~\hat{H}_1=\mathcal{A}\left(\hat{a}^2\hat{s}_K\hat{\sigma}_++\hat{a}^{\dagger 2}\hat{s}_K^\dagger\hat{\sigma}_-\right), \label{HNLone}\\
		\Delta_2&=0,~\hat{H}_2=\mathcal{B}\left(\hat{a}^2\hat{s}_K^\dagger\hat{\sigma}_++\hat{a}^{\dagger 2}\hat{s}_K\hat{\sigma}_-\right), \label{HNLTwo}\\
		\Delta_3&=0,~\hat{H}_3=\mathcal{A}\left(\hat{a}^2\hat{s}_K^\dagger\hat{\sigma}_-+\hat{a}^{\dagger 2}\hat{s}_K\hat{\sigma}_+\right). \label{HNLthree}
	\end{align}
	\label{HNL}
\end{subequations}
Here the coupling strengths are redefined as $\mathcal{A}=9\mathcal{C}\epsilon_0^2/2$ and $\mathcal{B}=15\mathcal{C}\epsilon_0^2/2$.
To verify the validity of these effective Hamiltonians, the dynamical simulation results of the original Hamiltonian are compared with those of the effective Hamiltonians $\hat{H}_1$, $\hat{H}_2$, and $\hat{H}_3$, as shown in Figs.~2(d), 2(e), and 2(f).

\begin{table*}
	\caption{\label{HKEALL}
		\textbf{Expansion of the Hamiltonian Eq.~(\ref{HKEThreeBody}).}
	}
	\begin{ruledtabular}
		\begin{tabular}{lccc}
			\textrm{No.}&
			\textrm{$g_x^{(2)}$}&
			\textrm{$g_y^{(2)}$}&
			\textrm{Frequency}\\
			\colrule
			1 & $\hat{a}^2\hat{s}_K\hat{\sigma}_+e^{-i(2\omega_z+\omega_K-\omega_s)}$ & $\hat{a}^2\hat{s}_K\hat{\sigma}_+e^{-i(2\omega_z+\omega_K-\omega_s)}$ & \textcolor{red}{$2\omega_z+\omega_K-\omega_s$}\\
			2 & $\hat{a}^2\hat{s}_K\hat{\sigma}_-e^{-i(2\omega_z+\omega_K+\omega_s)}$ & $-\hat{a}^2\hat{s}_K\hat{\sigma}_-e^{-i(2\omega_z+\omega_K+\omega_s)}$ & $2\omega_z+\omega_K+\omega_s$\\
			3 & $\hat{a}^2\hat{s}_K^\dagger\hat{\sigma}_+e^{-i(2\omega_z-\omega_K-\omega_s)}$ & $-\hat{a}^2\hat{s}_K^\dagger\hat{\sigma}_+e^{-i(2\omega_z-\omega_K-\omega_s)}$ & \textcolor{green}{$2\omega_z-\omega_K-\omega_s$}\\
			4 & $\hat{a}^2\hat{s}_K^\dagger\hat{\sigma}_-e^{-i(2\omega_z-\omega_K+\omega_s)}$ & $\hat{a}^2\hat{s}_K^\dagger\hat{\sigma}_-e^{-i(2\omega_z-\omega_K+\omega_s)}$ & \textcolor{blue}{$2\omega_z-\omega_K+\omega_s$}\\
			5 & $(\hat{a}^\dagger)^2\hat{s}_K\hat{\sigma}_+e^{i(2\omega_z-\omega_K+\omega_s)}$ & $(\hat{a}^\dagger)^2\hat{s}_K\hat{\sigma}_+e^{i(2\omega_z-\omega_K+\omega_s)}$ & \textcolor{blue}{$2\omega_z-\omega_K+\omega_s$}\\
			6 & $(\hat{a}^\dagger)^2\hat{s}_K\hat{\sigma}_-e^{i(2\omega_z-\omega_K-\omega_s)}$ & $-(\hat{a}^\dagger)^2\hat{s}_K\hat{\sigma}_-e^{i(2\omega_z-\omega_K-\omega_s)}$ & \textcolor{green}{$2\omega_z-\omega_K-\omega_s$}\\
			7 & $(\hat{a}^\dagger)^2\hat{s}_K^\dagger\hat{\sigma}_+e^{i(2\omega_z+\omega_K+\omega_s)}$ & $-(\hat{a}^\dagger)^2\hat{s}_K^\dagger\hat{\sigma}_+e^{i(2\omega_z+\omega_K+\omega_s)}$ & $2\omega_z+\omega_K+\omega_s$\\
			8 & $(\hat{a}^\dagger)^2\hat{s}_K^\dagger\hat{\sigma}_-e^{i(2\omega_z+\omega_K-\omega_s)}$ & $(\hat{a}^\dagger)^2\hat{s}_K^\dagger\hat{\sigma}_-e^{i(2\omega_z+\omega_K-\omega_s)}$ & \textcolor{red}{$2\omega_z+\omega_K-\omega_s$}
		\end{tabular}
	\end{ruledtabular}
\end{table*}

\subsection{\label{IMYD}Impact of motion in the $y$-direction}
The electron's motion along the $y$-axis is confined by a harmonic potential, characterized by the Hamiltonian
\begin{equation}
	\hat{H}_y = \frac{\hat{p}_y^2}{2m_e} + \frac{1}{2}m_e \omega_y^2 \hat{y}^2,
\end{equation}
where $\omega_y$ denotes the resonance frequency of the electron's motion along the $y$-axis.
The coupling between magnons and the electron's motion along the $y$-axis arises from magnetic dipole interactions. The quantized magnetic field acting on the electrons is given by
\begin{subequations}
	\begin{align}
		\hat{B}_x &= \frac{\mu_0 V_K M_K}{4\pi} \left(\frac{3H^2}{r^5} - \frac{1}{r^3}\right)\hat{m}_x,\\
		\hat{B}_y &= \frac{\mu_0 V_K M_K}{4\pi} \left(-\frac{1}{r^3}\right)\hat{m}_y,\\
		\hat{B}_z &= \frac{\mu_0 V_K M_K}{4\pi} \left(\frac{-3H z}{r^5}\right)\hat{m}_x
	\end{align}
\end{subequations}
with $r = \sqrt{z^2 + y^2 + H^2}$.
For $z/H,y/H\ll1$, expanding the magnetic field to second order yields
\begin{subequations}
	\begin{align}
		\hat{B}_x &= B_c \left(2 - \frac{6z^2}{H^2} - \frac{6y^2}{H^2} + \frac{45 y^2 \hat{z}^2}{2H^4}\right)\hat{m}_x,\\
		\hat{B}_y &= B_c \left(-1 + \frac{3z^2}{2H^2} +\frac{3y^2}{2H^2} - \frac{15y^2z^2}{4H^4}\right)\hat{m}_y,\\
		\hat{B}_z &= B_c \left(-\frac{3z}{H} + \frac{15y^2 z}{2 H^3}\right)\hat{m}_x.
	\end{align}
\end{subequations}
The quantized magnetic field induces coupling between the magnon and the electron's motion along the $z$-axis, coupling between the magnon and the $y$-axis motion, as well as coupling between the electron's motions along the $y$- and $z$-axes.
The coupling between the magnon and the electron's motion along the $z$-axis has been addressed in Section~\ref{TIH}.
We next discuss the coupling between the magnon and the electron's motion along the $y$-axis, as well as the coupling between the electron's motions along the $y$- and $z$-axes.

First, the coupling between the magnon and the electron's motion along the $y$-axis is described by the Hamiltonian
\begin{equation}
	\hat{H}_\mathrm{KE}^y=-\frac{\gamma_e}{2}\hat{\boldsymbol{B}}\cdot\boldsymbol{\sigma} = -\frac{\gamma_e}{2}\left(\hat{B}_x\hat{\sigma}_x + \hat{B}_y\hat{\sigma}_y + \hat{B}_z\hat{\sigma}_z\right).
\end{equation}
Incorporating the quantized magnetic field into the interaction Hamiltonian yields
\begin{equation}
	\hat{H}_\mathrm{KE}^y = -\frac{\gamma_eB_c}{2}\left(-\frac{6\hat{y}^2}{H^2}\hat{m}_x\hat{\sigma}_x + \frac{3\hat{y}^2}{2H^2}\hat{m}_y\hat{\sigma}_y\right).
\end{equation}
Introducing the quantized position coordinate $\hat{y} = y_0(\hat{c} + \hat{c}^\dagger)$ simplifies the interaction Hamiltonian to
\begin{equation}
	\hat{H}_\mathrm{KE}^y = \bar{g}_x^{(2)}\left(\hat{c} + \hat{c}^\dagger\right)^2\left(\hat{s}_K + \hat{s}_K^\dagger\right)\left(\hat{\sigma}_+ + \hat{\sigma}_-\right) - \bar{g}_y^{(2)}\left(\hat{c} + \hat{c}^\dagger\right)^2\left(\hat{s}_K - \hat{s}_K^\dagger\right)\left(\hat{\sigma}_+ - \hat{\sigma}_-\right),
\end{equation}
where the coupling strengths are given by $\bar{g}_x^{(2)}=6\mathcal{C}\epsilon_y^2$ and $\bar{g}_y^{(2)}=3\mathcal{C}\epsilon_y^2/2$, with $\epsilon_y=y_0/H$.
The coupling between the magnon and the electron's motion along the $y$-axis, and that between the magnon and the motion along the $z$-axis, exhibit similar forms.
Similarly, under the resonance conditions $\Delta_1^y=2\omega_y + \omega_K - \omega_s$, $\Delta_2^y=2\omega_y - \omega_K - \omega_s$, and $\Delta_3^y=2\omega_y - \omega_K + \omega_s$, various nonlinear interactions can be derived, analogous to those in Eq.~(\ref{HNL}).
We consider the case where the electron's motion along the $y$-axis is far off-resonant with respect to that along the $z$-axis.
Consequently, the resonance conditions ​$\Delta_{1,2,3}^y$ and $\Delta_{1,2,3}$ cannot be simultaneously satisfied due to the far-off-resonant between the electron's motion along the $y$- and $z$-axes.
Therefore, under the RWA, the coupling between the magnon and the electron's $y$-axis motion can be safely neglected due to the strong off-resonance between the $y$- and $z$-axis motions.

Next, we consider the coupling between the electron’s motion along the $y$-axis and that along the $z$-axis, described by the Hamiltonian
\begin{equation}
	\hat{H}_\mathrm{KE}^\mathrm{zy}=-\frac{\gamma_e}{2}\hat{\boldsymbol{B}}\cdot\boldsymbol{\sigma} = -\frac{\gamma_e}{2}\left(\hat{B}_x\hat{\sigma}_x + \hat{B}_y\hat{\sigma}_y + \hat{B}_z\hat{\sigma}_z\right).
\end{equation}
Substituting the quantized magnetic field into the interaction Hamiltonian yields
\begin{equation}
	\hat{H}_\mathrm{KE}^\mathrm{zy}=-\frac{\gamma_e B_c}{2}\left(\frac{45 \hat{y}^2 \hat{z}^2}{2H^4}\hat{m}_x\hat{\sigma}_x - \frac{15 \hat{y}^2 \hat{z}^2}{4H^4}\hat{m}_y\hat{\sigma}_y + \frac{15 \hat{y}^2 \hat{z}}{2H^3}\hat{m}_x\hat{\sigma}_z\right).
\end{equation}
Introducing the quantized displacement operators $\hat{y}=y_0(\hat{c}+\hat{c}^\dagger)$ and $\hat{z}=z_0(\hat{a}+\hat{a}^\dagger)$ simplifies the interaction Hamiltonian to
\begin{equation}
	\begin{split}
		\hat{H}_\mathrm{KE}^\mathrm{zy}=&\left(\hat{c}+\hat{c}^\dagger\right)^2\left(\hat{a}+\hat{a}^\dagger\right)^2\left[-g_x^{(4)}\left(\hat{s}_K + \hat{s}_K^\dagger\right)\left(\hat{\sigma}_+ + \hat{\sigma}_-\right) + g_y^{(4)}\left(\hat{s}_K-\hat{s}_K^\dagger\right)\left(\hat{\sigma}_+ - \hat{\sigma}_-\right)\right]\\
		&-g_z^{(3)}\left(\hat{c}+\hat{c}^\dagger\right)^2\left(\hat{a}+\hat{a}^\dagger\right)\hat{\sigma}_z,
	\end{split}
\end{equation}
where the coupling strengths are given by $g_x^{(4)}=45\mathcal{C}\epsilon_0^2\epsilon_y^2/2$,  $g_y^{(4)}=15\mathcal{C}\epsilon_0^2\epsilon_y^2/4$, and $g_z^{(3)}=15\mathcal{C}\epsilon_0\epsilon_y^2/2$.
On the one hand, since the electron’s motions along the $y$- and $z$-axes are far off-resonant, the coupling term $\hat{H}_\mathrm{KE}^\mathrm{zy}$ between these motions can be safely neglected.
On the other hand, the coupling strengths $g_x^{(4)}$, $g_y^{(4)}$, and $g_z^{(3)}$ correspond to third- and fourth-order small quantities ($\epsilon_0\epsilon_y^2$ and $\epsilon_0^2\epsilon_y^2$), and thus they are even smaller.
In summary, the Hamiltonian $\hat{H}_\mathrm{KE}^\mathrm{zy}$ can be safely neglected for the physical processes under consideration.

\section{\label{EECS}Exponentially enhanced coupling strength}
\subsection{\label{DoTH}Derivation of the Hamiltonian}
In the previous discussion, we provided a detailed analysis of the various types of nonlinear tripartite interactions that can be realized by selecting different resonance conditions.
Next, we will discuss methods for enhancing the coupling strength of nonlinear tripartite interactions.
The Hamiltonian of the electron's orbital state is given by
\begin{equation}
	\hat{H}_z=\frac{\hat{p}_z^2}{2m_e}+\frac{1}{2}k\hat{z}^2,
\end{equation}
where the stiffness coefficient is defined as $k=eV_0/(2W^2)$ with $V_0=V_c\exp(-2\pi d / W)$.
The stiffness coefficient can be further simplified to
\begin{equation}
	k=\frac{eV_c}{2W^2}\exp\left(-\frac{2\pi d}{W}\right).
\end{equation}
Consider a time-dependent modulation of the voltage $V(t)$, i.e.,
\begin{equation}
	V\left(t\right)=V_c-V_t\cos\left(2\omega_p t\right),
\end{equation}
which induces a time-dependent stiffness coefficient
\begin{equation}
	k\left(t\right)=k-k_t\cos\left(2\omega_d t\right),
\end{equation}
where
\begin{equation}
	k=\frac{eV_c}{2W^2}\exp\left(-\frac{2\pi d}{W}\right),~k_t=\frac{eV_t}{2W^2}\exp\left(-\frac{2\pi d}{W}\right).
\end{equation}
The Hamiltonian of the electron's orbital state can then be simplified to
\begin{equation}
	\hat{H}_z^\mathrm{NL}=\frac{\hat{p}_z^2}{2m_e}+\frac{1}{2}k\hat{z}^2-\frac{1}{2}k_t\cos\left(2\omega_p t\right)\hat{z}^2.
\end{equation}
Introducing the creation operator $\hat{a}^\dagger$ and the annihilation operator $\hat{a}$ for the electron's orbital state, the positional coordinates can be quantized as $\hat{z}=z_0(\hat{a}+\hat{a}^\dagger)$.
The Hamiltonian of the electron's orbital state, after the time-dependent modulation of the voltage, can be written as
\begin{equation}
	\hat{H}_z^\mathrm{NL}=\omega_z\hat{a}^\dagger\hat{a}-\Omega_p\cos\left(2\omega_p t\right)\left(\hat{a}+\hat{a}^\dagger\right)^2.
\end{equation}
That is, a nonlinear drive is applied to the orbital state through linear modulation of the voltage, with the drive strength $\Omega_p=k_tz_0^2/2$.

A detailed analysis of how this nonlinear drive can be utilized to achieve an exponential enhancement of the strength of the nonlinear tripartite interaction will follow.
Taking $\Delta_2=2\omega_z-\omega_K-\omega_s$ as an example, the Hamiltonian of the nonlinear tripartite interaction is given by
\begin{equation}
	\hat{H}_\mathrm{NKE}=\omega_z\hat{a}^\dagger\hat{a}+\omega_K\hat{s}_K^\dagger\hat{s}_K+\frac{\omega_s}{2}\hat{\sigma}_z+\mathcal{B}\left[\hat{a}^2\hat{s}_K^\dagger\hat{\sigma}_++\left(\hat{a}^\dagger\right)^2\hat{s}_K\hat{\sigma}_-\right]-\Omega_p\cos\left(2\omega_p t\right)\left(\hat{a}+\hat{a}^\dagger\right)^2.
\end{equation}
Transforming to a rotating frame with frequency $\omega_p$ and applying the rotating-wave approximation, we obtain the Hamiltonian
\begin{equation}
	\hat{H}_\mathrm{NKE}=\Delta_z\hat{a}^\dagger\hat{a}+\mathcal{W}_K\hat{s}_K^\dagger\hat{s}_K+\frac{\mathcal{W}_s}{2}\hat{\sigma}_z+\mathcal{B}(\hat{a}^2\hat{s}_K^\dagger\hat{\sigma}_+ + \hat{a}^{\dagger 2}\hat{s}_K\hat{\sigma}_-)-\frac{\Omega_p}{2}(\hat{a}^2+\hat{a}^{\dagger 2}),
\end{equation}
where the detunings are defined as $\Delta_z=\omega_z-\omega_p$ and $\mathcal{W}_{K,s}=\omega_{K,s}-\omega_p$.
Using the Bogoliubov transform $\hat{b}=\hat{a}\cosh r-\hat{a}^\dagger\sinh r$, the effective Hamiltonian of the system is given by
\begin{equation}
	\hat{H}_\mathrm{NKE}^\mathrm{eff}=\mathcal{W}_b\hat{b}^\dagger\hat{b}+\mathcal{W}_K\hat{s}_K^\dagger\hat{s}_K+\frac{\mathcal{W}_s}{2}\hat{\sigma}_z+\widetilde{\mathcal{B}}\left(\hat{b}^2\hat{s}_K^\dagger\hat{\sigma}_+ + \hat{b}^{\dagger 2}\hat{s}_K\hat{\sigma}_-\right),
\end{equation}
where the effective frequency is $\mathcal{W}_b=\Delta_z/\cosh(2r)$, the effective coupling strength is $\widetilde{\mathcal{B}}=\mathcal{B}\cosh^2 r$, and the squeezing parameter is $\tanh 2r=\Omega_p/\Delta_z$.
When the squeezing parameter $r$ is large, the effective coupling strength is reduced to $\widetilde{\mathcal{B}}/\mathcal{B}\propto\exp(2r)$.
That is, an exponential enhancement of the coupling strength is achieved, and the enhancement exceeds the conventional $\exp(r)$.
\subsection{\label{SAET}Stability analysis of electron trapping}
\begin{figure}
	\centering
	\includegraphics[width=0.5\textwidth]{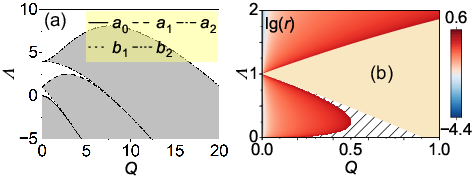}
	\caption{\textbf{Stability analysis of trapped electrons}. (a) Phase diagram of the Mathieu equation stability, with the gray regions indicating unstable parameter regimes. (b) Distribution of the squeezing parameter $r$ within the stable regions. The yellow-shaded region indicates the unstable regime, where the electron cannot be stably confined with two-phonon driving. The hatched area corresponds to nonphysical parameter values, where the parameter $\tanh(2r)$ exceeds unity.}
	\label{R_Fig1}
\end{figure}

We consider the motion of an electron in a potential well, for which the Hamiltonian in the absence of modulation is given by
\begin{equation}
	H_z=\frac{p_z^2}{2m_e}+\frac{1}{2}kz^2.
\end{equation}
The corresponding equation of motion is given by
\begin{equation}
	\ddot{z}+\frac{k}{m_e}z=0.
\end{equation}
When the potential well is periodically modulated, the spring constant becomes time-dependent and can be written as
\begin{equation}
	k(t)=k-k_t\cos(2\omega_p t).
\end{equation}
The equation of motion then becomes
\begin{equation}
	\ddot{z}+\frac{1}{m_e}\left[k-k_t\cos(2\omega_pt)\right]z=0.
\end{equation}
By introducing the dimensionless time $\mathcal{T}=\omega_p t$ and defining
\begin{equation}
	\Lambda=\frac{k}{m_e\omega_p^2}=\frac{\omega_z^2}{\omega_p^2},~Q=\frac{k_t}{2m_e\omega_p^2},~\omega_z=\sqrt{\frac{k}{m_e}}.
\end{equation}
This leads to the standard Mathieu equation
\begin{equation}
	\frac{d^2 z}{d\mathcal{T}^2}+\left[\Lambda-2Q\cos(2\mathcal{T})\right]=0.
\end{equation}
The stability of the solutions to this equation determines whether the electron can be stably trapped, i.e., whether its oscillation amplitude remains bounded.
In Fig.~\ref{R_Fig1}(a), a phase diagram is presented, showing the stable regions (white) and unstable regions (gray).
Fig.~\ref{R_Fig1}(b) further illustrates the distribution of $r$ within the stable regions, indicating that, with appropriate parameter choices, the squeezing parameter can reach $r=10^{0.6}\approx4$.

To further illustrate the feasibility of the experimental parameters, Table~\ref{Sqr_Para} lists several parameter sets within the stable regions, along with their corresponding squeezing parameters $r$.
The squeezing parameter is determined by $\tanh (2r)=\Omega_p/\Delta_z$, which requires precise control of the ratio $\Omega_p/\Delta_z$ to achieve a large squeezing parameter $r$.
Therefore, the feasibility of the scheme primarily depends on the precision of parameter tuning.
For the representative parameter sets above, we calculated the effective coupling strength $\widetilde{\mathcal{B}}$ and the amplified dissipation due to parametric driving, $\gamma_\mathrm{ph}^\mathrm{amp}=\gamma_\mathrm{ph} e^{2r}$.
The results are summarized in Table~\ref{Amp_Coupling_Strength}.
It is evident that, within the considered parameter range, the effective coupling strengths significantly exceed the corresponding dissipation rates, satisfying the strong-coupling condition.
Moreover, the system can enter the strong-coupling regime even for a moderate squeezing parameter of $r=2.6$

\begin{table*}
	\caption{\label{Sqr_Para}
		\textbf{Representative parameter sets within the stable regions and the corresponding squeezing parameters} $r$.
	}
	\begin{ruledtabular}
		\begin{tabular}{cccc}
			\textrm{$\Lambda$}&
			\textrm{$Q$}&
			\textrm{$\tanh (2r)$}&
			\textrm{$r$}\\
			\colrule
			0.01259 & 0.19942 & 0.99991 & 2.5\\
			0.12603 & 0.45845 & 0.99994 & 2.6\\
			0.128616 & 0.460546 & 0.999988 & 3.0\\
			0.4564016 & 0.4388439 & 0.9999988 & 3.6
		\end{tabular}
	\end{ruledtabular}
\end{table*}

\begin{table*}
	\caption{\label{Amp_Coupling_Strength}
		\textbf{Effective coupling strengths and the dissipation rates for representative squeezing parameters} $r$.
	}
	\begin{ruledtabular}
		\begin{tabular}{ccccc}
			\textrm{$r$}&
			\textrm{$\widetilde{\mathcal{B}}/2\pi~(\mathrm{MHz})$}&
			\textrm{$\gamma_\mathrm{ph}^\mathrm{amp}/2\pi~(\mathrm{kHz})$}&
			\textrm{$\gamma_s/2\pi~(\mathrm{kHz})$}&
			\textrm{$\gamma_K/2\pi~(\mathrm{MHz})$}\\
			\colrule
			2.6 & 0.54 & 3.0 & 10 & 0.49\\
			3.0 & 1.19 & 6.7 & 10 & 0.49\\
			3.6 & 3.95 & 22.3 & 10 & 0.49
		\end{tabular}
	\end{ruledtabular}
\end{table*}

\section{\label{MEEI}Magnon-mediated electron-electron interaction}
This section provides a detailed analysis of the physical mechanism through which two electrons are coupled via a magnon mediator.
According to the analysis in Sec.~\ref{EECS}, and following parametric amplification, the effective Hamiltonian describing two electrons coupled to a YIG sphere can be expressed as
\begin{equation}
	\hat{H}_{\mathrm{EKE}}=\sum_{i=1,2}\mathcal{W}_b^{(i)}\hat{b}_i^\dagger\hat{b}_i+\mathcal{W}_K\hat{s}_K^\dagger\hat{s}_K+\sum_{i=1,2}\frac{\mathcal{W}_s^{(i)}}{2}\hat{\sigma}_z^{(i)}+\sum_{i=1,2}\widetilde{\mathcal{B}}^{(i)}\left[\hat{b}_i^2\hat{s}_K^\dagger\hat{\sigma}_+^{(i)}+(\hat{b}_i^\dagger)^2\hat{s}_K\hat{\sigma}_-^{(i)}\right].
\end{equation}
For simplicity, we set $\mathcal{W}_b^{(1)}=\mathcal{W}_b^{(2)}=\mathcal{W}_b$, $\widetilde{\mathcal{B}}^{(1)}=\widetilde{\mathcal{B}}^{(2)}=\widetilde{\mathcal{B}}$, and $\mathcal{W}_s^{(1)}=\mathcal{W}_s^{(2)}=\mathcal{W}_s$.
By applying the unitary transformation $\hat{U}=\exp[-i\mathcal{W}_s(\hat{b}_1^\dagger\hat{b}_1+\hat{b}_2^\dagger\hat{b}_2+\hat{s}_K^\dagger\hat{s}_K+1/2\hat{\sigma}_z^{(1)}+1/2\hat{\sigma}_z^{(2)})t]$, we transform the Hamiltonian into the rotating frame
\begin{equation}
	\hat{H}_{\mathrm{EKE}}=\mathcal{D}_b\left(\hat{b}_1^\dagger\hat{b}_1+\hat{b}_2^\dagger\hat{b}_2\right)+\mathcal{D}_K\hat{s}_K^\dagger\hat{s}_K+\widetilde{\mathcal{B}}\sum_{i=1,2}\left[\hat{b}_i^2\hat{s}_K^\dagger\hat{\sigma}_+^{(i)}+(\hat{b}_i^\dagger)^2\hat{s}_K\hat{\sigma}_-^{(i)}\right],
\end{equation}
where $\mathcal{D}_b=\mathcal{W}_b-\mathcal{W}_s$ and $\mathcal{D}_K=\mathcal{W}_K-\mathcal{W}_s$.
Based on the above Hamiltonian, one obtains the quantum Langevin equations
\begin{subequations}
	\begin{align}
		\dot{\hat{b}}_1&=\left(-i\mathcal{D}_b-\frac{\gamma_\mathrm{ph}^\mathrm{amp}}{2}\right)\hat{b}_1-i\widetilde{\mathcal{B}}\left(2\hat{b}_1^\dagger\hat{s}_K\hat{\sigma}_-^{(1)}\right)-\sqrt{\gamma_\mathrm{ph}^\mathrm{amp}}\hat{b}_{1,\mathrm{in}}, \label{QEL_Sk_a}\\
		\dot{\hat{b}}_2&=\left(-i\mathcal{D}_b-\frac{\gamma_\mathrm{ph}^\mathrm{amp}}{2}\right)\hat{b}_2-i\widetilde{\mathcal{B}}\left(2\hat{b}_2^\dagger\hat{s}_K\hat{\sigma}_-^{(2)}\right)-\sqrt{\gamma_\mathrm{ph}^\mathrm{amp}}\hat{b}_{2,\mathrm{in}}, \label{QEL_Sk_b}\\
		\dot{\hat{s}}_K&=\left(-i\mathcal{D}_K-\frac{\gamma_K}{2}\right)\hat{s}_K-i\widetilde{\mathcal{B}}\hat{b}_1^2\hat{\sigma}_+^{(1)}-i\widetilde{\mathcal{B}}\hat{b}_2^2\hat{\sigma}_+^{(2)}-\sqrt{\gamma_K}\hat{s}_{K,\mathrm{in}}, \label{QEL_Sk}\\
		\dot{\hat{\sigma}}_-^{(1)}&=-2\gamma_s\hat{\sigma}_-^{(1)}+i\widetilde{\mathcal{B}}\hat{b}_1^2\hat{s}_K^\dagger\hat{\sigma}_z^{(1)}-\sqrt{\gamma_\mathrm{in}}\hat{\sigma}_{-,\mathrm{in}}^{(1)}, \label{QEL_Sk_d}\\
		\dot{\hat{\sigma}}_-^{(2)}&=-2\gamma_s\hat{\sigma}_-^{(2)}+i\widetilde{\mathcal{B}}\hat{b}_2^2\hat{s}_K^\dagger\hat{\sigma}_z^{(2)}-\sqrt{\gamma_\mathrm{in}}\hat{\sigma}_{-,\mathrm{in}}^{(2)} \label{QEL_Sk_e},
	\end{align}
\end{subequations}
where $\hat{b}_{1,\mathrm{in}}$, $\hat{b}_{2,\mathrm{in}}$, $\hat{s}_{K,\mathrm{in}}$, $\hat{\sigma}_{-,\mathrm{in}}^{(1)}$, and $\hat{\sigma}_{-,\mathrm{in}}^{(2)}$ denote noise operators.
Here, $\gamma_\mathrm{in}$ denotes the external coupling rate.
The formal solutions of these equations are given by
\begin{subequations}
	\begin{align}
		\hat{b}_1(t)&=\hat{b}_1(0)\exp\left[\left(-i\mathcal{D}_b-\frac{\gamma_\mathrm{ph}^\mathrm{amp}}{2}\right)t\right] \nonumber \\
		&+\int_0^t\left[-i\widetilde{\mathcal{B}}\left(2\hat{b}_1^\dagger(\tau)\hat{s}_K(\tau)\hat{\sigma}_-^{(1)}(\tau)\right)-\sqrt{\gamma_\mathrm{ph}^\mathrm{amp}}\hat{b}_{1,\mathrm{in}}(\tau)\right]\exp\left[\left(-i\mathcal{D}_b-\frac{\gamma_\mathrm{ph}^\mathrm{amp}}{2}\right)\left(t-\tau\right)\right]d\tau,\\
		\hat{b}_2(t)&=\hat{b}_2(0)\exp\left[\left(-i\mathcal{D}_b-\frac{\gamma_\mathrm{ph}^\mathrm{amp}}{2}\right)t\right] \nonumber \\
		&+\int_0^t\left[-i\widetilde{\mathcal{B}}\left(2\hat{b}_2^\dagger(\tau)\hat{s}_K(\tau)\hat{\sigma}_-^{(2)}(\tau)\right)-\sqrt{\gamma_\mathrm{ph}^\mathrm{amp}}\hat{b}_{2,\mathrm{in}}(\tau)\right]\exp\left[\left(-i\mathcal{D}_b-\frac{\gamma_\mathrm{ph}^\mathrm{amp}}{2}\right)\left(t-\tau\right)\right]d\tau,\\
		\hat{s}_K(t)&=\hat{s}_K(0)\exp[\left(-i\mathcal{D}_K-\frac{\gamma_K}{2}\right)t] \nonumber \\
		&+\int_0^t\left[-i\widetilde{\mathcal{B}}\hat{b}_1^2(\tau)\hat{\sigma}_+^{(1)}(\tau)-i\widetilde{\mathcal{B}}\hat{b}_2^2(\tau)\hat{\sigma}_+^{(2)}(\tau)-\sqrt{\gamma_K}\hat{s}_{K,\mathrm{in}}(\tau)\right]\exp\left[\left(-i\mathcal{D}_K-\frac{\gamma_K}{2}\right)(t-\tau)\right]d\tau,\\
		\hat{\sigma}_-^{(1)}(t)&=\hat{\sigma}_-^{(1)}(0)\exp\left(-2\gamma_st\right)+\int_0^t \left[i\widetilde{\mathcal{B}}\hat{b}_1^2(\tau)\hat{s}_K^\dagger(\tau)\hat{\sigma}_z^{(1)}(\tau)-\sqrt{\gamma_\mathrm{in}}\hat{\sigma}_{-,\mathrm{in}}^{(1)}(\tau)\right]\exp\left[-2\gamma_s(t-\tau)\right]d\tau,\\				
		\hat{\sigma}_-^{(2)}(t)&=\hat{\sigma}_-^{(2)}(0)\exp\left(-2\gamma_st\right)+\int_0^t \left[i\widetilde{\mathcal{B}}\hat{b}_2^2(\tau)\hat{s}_K^\dagger(\tau)\hat{\sigma}_z^{(2)}(\tau)-\sqrt{\gamma_\mathrm{in}}\hat{\sigma}_{-,\mathrm{in}}^{(2)}(\tau)\right]\exp\left[-2\gamma_s(t-\tau)\right]d\tau.
	\end{align}
\end{subequations}
We focus on the regime $\vert \mathcal{D}_K\vert \gg \widetilde{\mathcal{B}}$, where the magnon mode is far detuned.
In this regime, the dynamics of $\hat{b}_{1,2}$ and $\hat{\sigma}_-^{(1,2)}$ are only weakly affected by the magnon mode, so that
\begin{subequations}
	\begin{align}
		\hat{b}_1(t)&=\hat{b}_1(0)\exp\left[\left(-i\mathcal{D}_b-\frac{\gamma_\mathrm{ph}^\mathrm{amp}}{2}\right)t\right] + \hat{B}_{i,\mathrm{in}}(t),\\
		\hat{b}_2(t)&=\hat{b}_2(0)\exp\left[\left(-i\mathcal{D}_b-\frac{\gamma_\mathrm{ph}^\mathrm{amp}}{2}\right)t\right] + \hat{B}_{2,\mathrm{in}}(t),\\
		\hat{\sigma}_-^{(1)}(t)&=\hat{\sigma}_-^{(1)}(0)\exp\left(-2\gamma_st\right) + \hat{\Sigma}_{-,\mathrm{in}}^{(1)}(t),\\	\hat{\sigma}_-^{(2)}(t)&=\hat{\sigma}_-^{(2)}(0)\exp\left(-2\gamma_st\right)+ \hat{\Sigma}_{-,\mathrm{in}}^{(2)}(t),
	\end{align}
	\label{QEL_Big_Detuned}
\end{subequations}
where $\hat{B}_{i,\mathrm{in}}(t)$, $\hat{B}_{2,\mathrm{in}}(t)$, $\hat{\Sigma}_{-,\mathrm{in}}^{(1)}(t)$, and $\hat{\Sigma}_{-,\mathrm{in}}^{(2)}(t)$ denote noise operators.
Substituting the Eqs.~(\ref{QEL_Big_Detuned}) into the equation of motion for the magnon mode [Eq.~(\ref{QEL_Sk})] and performing the integration yields
\begin{equation}
	\hat{s}_K(t)=\hat{s}_K(0)\exp\left[\left(-i\mathcal{D}_K-\frac{\gamma_K}{2}\right)t\right]+\frac{-i\widetilde{\mathcal{B}}\hat{b}_1^2(t)\hat{\sigma}_+^{(1)}(t)-i\widetilde{\mathcal{B}}\hat{b}_2^2(t)\hat{\sigma}_+^{(2)}(t)}{i\left(\mathcal{D}_K-2\mathcal{D}_b\right)+\frac{\Gamma}{2}}+\hat{S}_{K,\mathrm{in}}(t),
\end{equation}
where $\Gamma=\gamma_K-2\gamma_\mathrm{ph}^\mathrm{amp}-4\gamma_s$. In the parameter regime of interest, where $\mathcal{D}_K\gg 2\mathcal{D}_b$, and in the presence of strong magnon dissipation, the first term can be safely neglected, yielding
\begin{equation}
	\hat{s}_K(t)=\frac{-i\widetilde{\mathcal{B}}\hat{b}_1^2(t)\hat{\sigma}_+^{(1)}(t)-i\widetilde{\mathcal{B}}\hat{b}_2^2(t)\hat{\sigma}_+^{(2)}(t)}{i\left(\mathcal{D}_K-2\mathcal{D}_b\right)+\frac{\Gamma}{2}}+\hat{S}_{K,\mathrm{in}}(t),
	\label{QEL_sK_eff}		
\end{equation}
where $\hat{S}_{K,\mathrm{in}}(t)$ denotes noise operator.
Substituting the Eq.~(\ref{QEL_sK_eff}) back into the quantum Langevin equations [Eqs.~(\ref{QEL_Sk_a},~\ref{QEL_Sk_b},~\ref{QEL_Sk_d},~\ref{QEL_Sk_e})] and simplifying, we obtain
\begin{subequations}
	\begin{align}
		\dot{\hat{b}}_1&=\left(-i\mathcal{D}_b-\frac{\gamma_\mathrm{ph}^\mathrm{amp}}{2}\right)\hat{b}_1-i\left(\frac{\widetilde{B}^2 2\hat{b}_1^\dagger\hat{b}_1^2\hat{\sigma}_+^{(1)}\hat{\sigma}_-^{(1)}}{-\mathcal{D}_K+i\Gamma/2}+\frac{\widetilde{B}^2 2\hat{b}_1^\dagger\hat{b}_2^2\hat{\sigma}_+^{(2)}\hat{\sigma}_-^{(1)}}{-\mathcal{D}_K+i\Gamma/2}\right)+\hat{\widetilde{B}}_{1,\mathrm{in}},\\
		\dot{\hat{b}}_2&=\left(-i\mathcal{D}_b-\frac{\gamma_\mathrm{ph}^\mathrm{amp}}{2}\right)\hat{b}_2-i\left(\frac{\widetilde{B}^2 2\hat{b}_2^\dagger\hat{b}_1^2\hat{\sigma}_+^{(1)}\hat{\sigma}_-^{(2)}}{-\mathcal{D}_K+i\Gamma/2}+\frac{\widetilde{B}^2 2\hat{b}_2^\dagger\hat{b}_2^2\hat{\sigma}_+^{(2)}\hat{\sigma}_-^{(2)}}{-\mathcal{D}_K+i\Gamma/2}\right)+\hat{\widetilde{B}}_{2,\mathrm{in}},\\
		\dot{\hat{\sigma}}_-^{(1)}&=-2\gamma_s\hat{\sigma}_-^{(1)}-i\left(\frac{\widetilde{B}^2 \hat{b}_1^2(\hat{b}_1^\dagger)^2\hat{\sigma}_-^{(1)}}{\mathcal{D}_K+i\Gamma/2}+\frac{\widetilde{B}^2 \hat{b}_1^2(\hat{b}_2^\dagger)^2\hat{\sigma}_-^{(2)}\hat{\sigma}_z^{(1)}}{\mathcal{D}_K+i\Gamma/2}\right)+\hat{\widetilde{\Sigma}}_{-,\mathrm{in}}^{(1)},\\
		\dot{\hat{\sigma}}_-^{(2)}&=-2\gamma_s\hat{\sigma}_-^{(2)}-i\left(\frac{\widetilde{B}^2 \hat{b}_2^2(\hat{b}_1^\dagger)^2\hat{\sigma}_-^{(1)}\hat{\sigma}_z^{(2)}}{\mathcal{D}_K+i\Gamma/2}+\frac{\widetilde{B}^2 \hat{b}_2^2(\hat{b}_2^\dagger)^2\hat{\sigma}_-^{(2)}}{\mathcal{D}_K+i\Gamma/2}\right)+\hat{\widetilde{\Sigma}}_{-,\mathrm{in}}^{(2)}.
	\end{align}
	\label{QEL_Without_Sk}
\end{subequations}
Here, $\hat{\widetilde{B}}_{1,\mathrm{in}}$, $\hat{\widetilde{B}}_{2,\mathrm{in}}$, $\hat{\widetilde{\Sigma}}_{-,\mathrm{in}}^{(1)}$, and $\hat{\widetilde{\Sigma}}_{-,\mathrm{in}}^{(2)}$ denote the corresponding modified noise operators.
The effective Hamiltonian corresponding to Eqs.~(\ref{QEL_Without_Sk}) is given by
\begin{equation}
	\begin{split}
		\hat{H}_\mathrm{EE}&=\mathcal{D}_b\left(\hat{b}_1^\dagger\hat{b}_1+\hat{b}_2^\dagger\hat{b}_2\right)+\frac{1}{2}\left[\frac{\widetilde{\mathcal{B}}^2}{-\mathcal{D}_K+i\Gamma/2}\left((\hat{b}_1^\dagger)^2\hat{b}_1^2\hat{\sigma}_+^{(1)}\hat{\sigma}_-^{(1)}+(\hat{b}_2^\dagger)^2\hat{b}_2^2\hat{\sigma}_+^{(2)}\hat{\sigma}_-^{(2)}\right)\right]\\
		&+\frac{1}{2}\left[\frac{\widetilde{\mathcal{B}}^2}{\mathcal{D}_K+i\Gamma/2}\left(\hat{b}_1^2(\hat{b}_1^\dagger)^2\frac{\hat{\sigma}_z^{(1)}}{2}+\hat{b}_2^2(\hat{b}_2^\dagger)^2\frac{\hat{\sigma}_z^{(2)}}{2}\right)\right]\\
		&+\frac{\widetilde{\mathcal{B}}^2}{2}\left(\frac{1}{-\mathcal{D}_K+i\Gamma/2}-\frac{1}{\mathcal{D}_K+i\Gamma/2}\right)\left((\hat{b}_1^\dagger)^2\hat{b}_2^2\hat{\sigma}_-^{(1)}\hat{\sigma}_+^{(2)}+\hat{b}_1^2(\hat{b}_2^\dagger)^2\hat{\sigma}_+^{(1)}\hat{\sigma}_-^{(2)}\right).
	\end{split}
\end{equation}
Here, the factor $1/2$ outside the brackets in the last three terms is introduced to avoid double counting of the coupling channels.
Using the identity $\hat{\sigma}_+\hat{\sigma}_-=(\mathbb{I}+\hat{\sigma}_z)/2$ (Here, $\mathbb{I}$ denotes the identity matrix), the Hamiltonian can be simplified as follows
\begin{equation}
	\begin{split}
		\hat{H}_\mathrm{EE}&=\mathcal{D}_b\left(\hat{b}_1^\dagger\hat{b}_1+\hat{b}_2^\dagger\hat{b}_2\right)-\frac{\widetilde{\mathcal{B}}^2}{4}\frac{\mathcal{D}_K+i\Gamma/2}{\mathcal{D}_K^2+\Gamma^2/4}\left((\hat{b}_1^\dagger)^2\hat{b}_1^2+(\hat{b}_2^\dagger)^2\hat{b}_2^2\right)\\
		&-\frac{\widetilde{\mathcal{B}}^2}{2}\frac{\mathcal{D}_K+i\Gamma}{\mathcal{D}_K^2+\Gamma^2/4}\left((\hat{b}_1^\dagger)^2\hat{b}_1^2\frac{\hat{\sigma}_z^{(1)}}{2}+(\hat{b}_2^\dagger)^2\hat{b}_2^2\frac{\hat{\sigma}_z^{(2)}}{2}\right)\\
		&-\frac{\widetilde{\mathcal{B}}^2}{2}\frac{-\mathcal{D}_K+i\Gamma}{\mathcal{D}_K^2+\Gamma^2/4}\left((\hat{b}_1^2\hat{b}_1^\dagger)^2\frac{\hat{\sigma}_z^{(1)}}{2}+\hat{b}_2^2(\hat{b}_2^\dagger)^2\frac{\hat{\sigma}_z^{(2)}}{2}\right)\\
		&-\frac{\widetilde{\mathcal{B}}^2\mathcal{D}_K}{\mathcal{D}_K^2+\Gamma^2/4}\left((\hat{b}_1^\dagger)^2\hat{b}_2^2\hat{\sigma}_-^{(1)}\hat{\sigma}_+^{(2)}+\hat{b}_1^2(\hat{b}_2^\dagger)^2\hat{\sigma}_+^{(1)}\hat{\sigma}_-^{(2)}\right).
	\end{split}
\end{equation}
In the considered regime $\mathcal{D}_K \gg (\Gamma,~\gamma_K)$, the small term $\Gamma/\mathcal{D}_K^2$ can be safely neglected, yielding
\begin{equation}
	\begin{split}
		\hat{H}_\mathrm{EE}&=\mathcal{D}_b\left(\hat{b}_1^\dagger\hat{b}_1+\hat{b}_2^\dagger\hat{b}_2\right)-\mathcal{K}_\mathrm{EE}\left((\hat{b}_1^\dagger)^2\hat{b}_1^2+(\hat{b}_2^\dagger)^2\hat{b}_2^2\right)	\\	
		&+\mathcal{K}_\mathrm{ES}\left[\left(2\hat{b}_1^\dagger\hat{b}_1+1\right)\hat{\sigma}_z^{(1)}+\left(2\hat{b}_2^\dagger\hat{b}_2+1\right)\hat{\sigma}_z^{(2)}\right]\\
		&-\mathcal{G}_\mathrm{EE}\left((\hat{b}_1^\dagger)^2\hat{b}_2^2\hat{\sigma}_-^{(1)}\hat{\sigma}_+^{(2)}+\hat{b}_1^2(\hat{b}_2^\dagger)^2\hat{\sigma}_+^{(1)}\hat{\sigma}_-^{(2)}\right),
	\end{split}
\end{equation}
where $\mathcal{K}_\mathrm{EE}=\widetilde{\mathcal{B}}^2/(4\mathcal{D}_K)$, $\mathcal{K}_\mathrm{ES}=\widetilde{\mathcal{B}}^2/(2\mathcal{D}_K)$, and $\mathcal{G}_\mathrm{EE}=\widetilde{\mathcal{B}}^2/\mathcal{D}_K$.
We consider the regime $\mathcal{D}_b \gg \mathcal{K}_\mathrm{EE},~\mathcal{K}_\mathrm{ES}$, where the corrections $[(\hat{b}_1^\dagger)^2\hat{b}_1^2+(\hat{b}_2^\dagger)^2\hat{b}_2^2]$ and $[(2\hat{b}_1^\dagger\hat{b}_1+1)\hat{\sigma}_z^{(1)}+(2\hat{b}_2^\dagger\hat{b}_2+1)\hat{\sigma}_z^{(2)}]$ have a negligible effect on the oscillator energies, resulting in
\begin{equation}
	\hat{H}_\mathrm{EE}=\mathcal{D}_b\left(\hat{b}_1^\dagger\hat{b}_1+\hat{b}_2^\dagger\hat{b}_2\right)-\mathcal{G}_\mathrm{EE}\left((\hat{b}_1^\dagger)^2\hat{b}_2^2\hat{\sigma}_-^{(1)}\hat{\sigma}_+^{(2)}+\hat{b}_1^2(\hat{b}_2^\dagger)^2\hat{\sigma}_+^{(1)}\hat{\sigma}_-^{(2)}\right).
\end{equation}
Transforming back to the original representation, we have
\begin{equation}
	\hat{H}_\mathrm{EE}=\mathcal{W}_b\left(\hat{b}_1^\dagger\hat{b}_1+\hat{b}_2^\dagger\hat{b}_2\right)+\frac{\mathcal{W}_s}{2}\left(\hat{\sigma}_z^{(1)}+\hat{\sigma}_z^{(2)}\right)-\mathcal{G}_\mathrm{EE}\left((\hat{b}_1^\dagger)^2\hat{b}_2^2\hat{\sigma}_-^{(1)}\hat{\sigma}_+^{(2)}+\hat{b}_1^2(\hat{b}_2^\dagger)^2\hat{\sigma}_+^{(1)}\hat{\sigma}_-^{(2)}\right).
\end{equation}
Transforming into the interaction picture, we arrive at the effective electron-electron interaction Hamiltonian
\begin{equation}
	\hat{H}_\mathrm{EE}=-\mathcal{G}_\mathrm{EE}\left((\hat{b}_1^\dagger)^2\hat{b}_2^2\hat{\sigma}_-^{(1)}\hat{\sigma}_+^{(2)}+\hat{b}_1^2(\hat{b}_2^\dagger)^2\hat{\sigma}_+^{(1)}\hat{\sigma}_-^{(2)}\right).
\end{equation}
From the complete derivation above, it is evident that magnon dissipation $\gamma_K$ enters the effective Hamiltonian only as a small correction of order $\Gamma/\mathcal{D}_K^2$.
Within the parameter regime of interest, $\mathcal{D}_K\gg (\widetilde{\mathcal{B}},~\Gamma,~\gamma_K)$, meaning that the magnon mode is far detuned and its dissipation is much smaller than the detuning, the effect of magnon dissipation on the system dynamics can be safely neglected.
Moreover, the parametric amplification mechanism enhances only the effective coupling strength of the phonon modes and does not directly increase the dissipation of the magnon mode.

\begin{figure}
	\centering
	\includegraphics[width=0.5\textwidth]{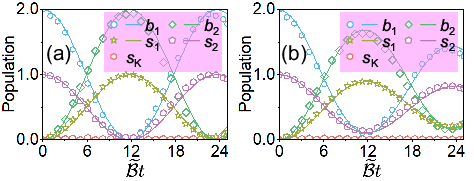}
	\caption{\textbf{Electron-electron interaction mediated by magnons.} Dynamics of the original Hamiltonian $\hat{H}_{\mathrm{EKE}}$ versus the effective Hamiltonian $\hat{H}_{\mathrm{EE}}$. Here, $b_i$ and $s_i$ denote the phonon and spin degrees of freedom of the $i$-th electron, respectively, while $s_K$ represents the Kittel mode. The parameters are set as $\mathcal{D}_K=15\widetilde{\mathcal{B}}$, $\mathcal{D}_b=10\mathcal{K}_\mathrm{ES}$, and $\mathcal{W}_s=10\widetilde{\mathcal{B}}$.}
	\label{R_Fig2}
\end{figure}

To validate the above conclusion, numerical simulations are conducted using the parameters summarized in Table~\ref{Origin_VS_EFF_only_magnon_diss}.
For the original Hamiltonian $\hat{H}_\mathrm{EKE}$, only magnon dissipation is taken into account.
For the effective Hamiltonian $\hat{H}_\mathrm{EE}$, magnon dissipation is neglected, as the magnon mode is absent.
For consistency, the dissipation of the trapped-electron system is also neglected.
Good agreement between the dynamical evolutions of the two Hamiltonians indicates that the effect of magnon dissipation can be safely neglected.
The numerical results are presented in Fig.~\ref{R_Fig2}(a).
The dynamical evolutions of the two Hamiltonians exhibit excellent agreement, indicating that magnon dissipation has a negligible effect on the magnon-mediated effective interaction between electrons and can therefore be safely ignored.
In Fig.~\ref{R_Fig2}(b), numerical simulations are performed using the parameters listed in Table~\ref{Origin_VS_EFF_All_diss}, with both magnon dissipation and the dissipation of the trapped-electron system taken into account.
The results again show excellent agreement, further confirming that neglecting magnon dissipation is a valid and well-justified approximation.

\begin{table*}
	\caption{\label{Origin_VS_EFF_only_magnon_diss}
		\textbf{Simulation Parameters for Fig.~\ref{R_Fig2}(a)}.
	}
	\begin{ruledtabular}
		\begin{tabular}{cccc}
			\textrm{Hamiltonian}&
			\textrm{$\gamma_K/\widetilde{\mathcal{B}}$}&
			\textrm{$\gamma_\mathrm{ph}^\mathrm{amp}/\widetilde{\mathcal{B}}$}&
			\textrm{$\gamma_s/\widetilde{\mathcal{B}}$}\\
			\colrule
			$\hat{H}_\mathrm{EKE}$ & 0.1 & 0 & 0\\
			$\hat{H}_\mathrm{EE}$ & None & 0 & 0
		\end{tabular}
	\end{ruledtabular}
\end{table*}

\begin{table*}
	\caption{\label{Origin_VS_EFF_All_diss}
		\textbf{Simulation Parameters for Fig.~\ref{R_Fig2}(b)}.
	}
	\begin{ruledtabular}
		\begin{tabular}{cccc}
			\textrm{Hamiltonian}&
			\textrm{$\gamma_K/\widetilde{\mathcal{B}}$}&
			\textrm{$\gamma_\mathrm{ph}^\mathrm{amp}/\widetilde{\mathcal{B}}$}&
			\textrm{$\gamma_s/\widetilde{\mathcal{B}}$}\\
			\colrule
			$\hat{H}_\mathrm{EKE}$ & 0.1 & 0.01 & 0.01\\
			$\hat{H}_\mathrm{EE}$ & None & 0.01 & 0.01
		\end{tabular}
	\end{ruledtabular}
\end{table*}

\section{\label{NGSP}Non-Gaussian state preparation}
\subsection{\label{MMDC}Magnon-mediated dissipative coupling}
In Sec.~\ref{MEEI}, we examined the indirect coherent coupling among the four degrees of freedom of two electrons mediated by magnons.
In this section, the magnon is further treated as a dissipative mediator, exploiting its strong dissipation to mediate dissipative coupling between the phonon and spin degrees of freedom within a single electron.
We analyze in detail how, under conditions of strong magnon dissipation, adiabatic elimination of the magnon mode can effectively generate a dissipative coupling channel between the phonon and spin degrees of freedom within a single electron.
The subsequent analysis focuses on the resonance condition $\Delta_1=0$.
Based on the theoretical derivation in Sec.~\ref{EECS}, the effective interaction Hamiltonian after parametric amplification can be written as
\begin{equation}
	\hat{\mathfrak{H}}_\mathrm{NKE}=\mathfrak{W}_z\hat{b}^\dagger\hat{b}+\mathfrak{W}_K\hat{s}_K^\dagger\hat{s}_K+\frac{\mathfrak{W}_s}{2}\hat{\sigma_z}+\widetilde{\mathcal{A}}\left[\hat{b}^2\hat{s}_K\hat{\sigma}_++\left(\hat{b}^\dagger\right)^2\hat{s}_K^\dagger\hat{\sigma}_-\right],
\end{equation}
where the detuning parameters are $\mathfrak{W}_z=\mathfrak{D}_z/\cosh(2r)$, $\mathfrak{D}_z=\omega_z-\omega_p/2$, $\mathfrak{W}_K=\omega_K - \omega_p$, and $\mathfrak{W}_s=\omega_s-2\omega_p$.
The exponentially enhanced coupling strength is given by $\widetilde{\mathcal{A}}=\mathcal{A}\cosh^2 r$.
Transforming to the interaction picture, we obtain
\begin{equation}
	\hat{\mathfrak{H}}_\mathrm{NKE}=\widetilde{\mathcal{A}}\left[\hat{b}^2\hat{s}_K\hat{\sigma}_++\left(\hat{b}^\dagger\right)^2\hat{s}_K^\dagger\hat{\sigma}_-\right].
\end{equation}
Considering the magnon dissipation $\gamma_K$, the master equation of the system can be written as
\begin{equation}
	\dot{\hat{\rho}}=-i\left[\hat{\mathfrak{H}}_\mathrm{NKE},\hat{\rho}\right]+\frac{\gamma_K}{2}\left(2\hat{s}_K\hat{\rho}\hat{s}_K^\dagger-\hat{s}_K^\dagger\hat{s}_K\hat{\rho}-\hat{\rho}\hat{s}_K^\dagger\hat{s}_K\right).
	\label{MSENLOne}
\end{equation}
The equation of motion for the magnon operator $\hat{s}_K$ is given by
\begin{equation}
	\partial_t\hat{s}_K=-i\widetilde{\mathcal{A}}\left(\hat{b}^\dagger\right)^2\hat{\sigma}_--\frac{\gamma_K}{2}\hat{s}_K+\sqrt{\gamma_K}\hat{s}_K^\mathrm{in},
\end{equation}
where $\hat{s}_K^\mathrm{in}$ represents the noise operator.
When the magnon dissipation is large, the magnon remains in the steady state $\partial_t\hat{s}_K=0$, leading to
\begin{equation}
	\hat{s}_K=-\frac{2i\widetilde{\mathcal{A}}}{\gamma_K}\left(\hat{b}^\dagger\right)^2\hat{\sigma}_-,~
	\hat{s}_K^\dagger=\frac{2i\widetilde{\mathcal{A}}}{\gamma_K}\hat{b}^2\hat{\sigma}_+.
	\label{EQSk}
\end{equation}
Based on Eq.~(\ref{EQSk}) and Eq.~(\ref{MSENLOne}), the effective master equation of the system is given by
\begin{equation}
	\dot{\hat{\rho}}=\frac{4\widetilde{\mathcal{A}}^2}{\gamma_K}D\left[\left(\hat{b}^\dagger\right)^2\hat{\sigma}_-\right].
	\label{D1EFF}
\end{equation}
This equation characterizes the nonlinear dissipative interaction between the electron's motional and spin states, where the annihilation of a single spin excitation results in the generation of two orbital excitations.

\subsection{Non-Gaussian state preparation.}
The preparation and control of nonclassical states are core to quantum optics technology~\cite{2004WengerP153601153601,2021WalschaersP3020430204,2023HuP3203132031}.
Non-Gaussian states can be prepared by adding single-quantum excitations to Gaussian states, which is an important approach for realizing non-Gaussian states, defined by $\vert \alpha,m\rangle=k_{\alpha,m}(\hat{c}^{\dagger})^m\vert \alpha\rangle$, where $\hat{c}^\dagger$ is the creation operator of an arbitrary boson, $m$ represents the number of times the creation operator acts, and $\alpha$ represents the displacement of the coherent state~\cite{1991AgarwalP492497,2005ZavattaP2382023820}.
For example, a photon-added coherent state is a novel quantum state generated by successive applications of the creation operator to the coherent state, which exhibits a negative Wigner function~\cite{2004ZavattaP660662,2005ZavattaP2382023820,2010BarbieriP6383363833,2024FadrnyP8989}.
\begin{figure}
	\centering
	\includegraphics[width=0.5\textwidth]{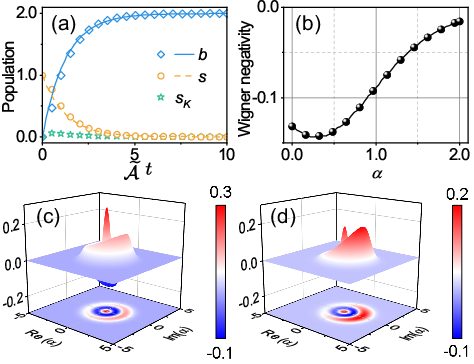}
	\caption{\textbf{Non-Gaussian State Preparation.} (a) Dynamical comparison between the original and effective master equations. (b) Wigner negativity versus initial coherent state size $\alpha$. (c) and (d) then show the Wigner function for the steady phonon-added coherent state of the system when the initial state is $\vert \alpha=0.3 \rangle$ and $\vert \alpha=0.7 \rangle$ respectively.}
	\label{SMFig3}
\end{figure}

In the following, we utilize dissipative interactions induced by the large dissipation of magnons to mimic spin-dependent phonon-added operations.
According to the analysis in Sec.\ref{MMDC}, when the magnon dissipation rate is much larger than the other characteristic rates of the system (i.e., in the large-dissipation limit), the system’s overall dynamics can be effectively described by the master equation (\ref{D1EFF}).
The master equation describes a dynamical process in which the annihilation of one spin excitation results in two motional state excitations, corresponding to the application of two phonon creation operations on the motional state.
It is important to note that this process of phonon-added operations depends on the initial spin state.
The phonon-added operation is conditionally executable in the spin's excited state but remains prohibited under ground-state spin configurations.
Figure~\ref{SMFig3}(a) shows that when the spin is in the excited state $\vert \uparrow\rangle$, the motional state gains two excitations as the system evolves, accompanied by spin relaxation to the ground state.
The scatter points and solid curves in the Fig.~\ref{SMFig3}(a) represent the simulation results obtained from the original [Eq.~(\ref{MSENLOne})] and effective [Eq.~(\ref{D1EFF})] master equations, respectively.
Their close agreement verifies the validity of the effective Hamiltonian.

We define Wigner negativity as a measure of the non-Gaussianity of the final steady state, given by the minimum value of the Wigner function $W$, i.e., $\mathcal{W}=\min\{W\}$~\cite{1932WignerP749759,1997GerryP964974,2023HauerP213604213604}.
Fig. \ref{SMFig3}(b) shows that the non-Gaussianity of the final steady state decreases as the size of the coherent state increases.
This can be attributed to the increase in the average phonon number as the size of the coherent state grows.
When the average phonon number is large, adding two phonons to the coherent state has little effect, resulting in a reduction of the non-Gaussianity of the steady state of the electron motional state.
Figs.~\ref{SMFig3}(c) and \ref{SMFig3}(d) show the Wigner function of the electron motional state at steady state for initial coherent states
$\vert \alpha=0.3\rangle$ and $\vert \alpha=0.7\rangle$, respectively, with the spin initialized in the $\vert \uparrow\rangle$.
The Wigner function of the electron motional state is negative, indicating that the steady state of the motional state is non-Gaussian.
The simulation results presented here are in agreement with experimental findings for photons~\cite{2024FadrnyP8989}.

\bibliography{SMReference}